\documentclass[preprint2]{aastex62}
\usepackage{graphicx}
\usepackage{xcolor}
\usepackage{epstopdf}
\usepackage{amsmath}
\usepackage{hyperref}
\usepackage{cleveref}
\usepackage{natbib}
\usepackage{lineno}

\hypersetup{linkcolor=red,citecolor=green,filecolor=cyan,urlcolor=magenta}

\def\simlt{\mathrel{\hbox{\rlap{\hbox{\lower4pt\hbox{$\sim$}}}\hbox{$<$}}}}
\def\simgt{\mathrel{\hbox{\rlap{\hbox{\lower4pt\hbox{$\sim$}}}\hbox{$>$}}}}

\usepackage[scr]{rsfso}

\usepackage[utf8]{inputenc}

\graphicspath{./}

\submitjournal{ApJL}

\shorttitle{He-rich SLSN-I from ZTF }
\shortauthors{Yan et al.}

\begin{document}
\title{Helium-rich Superluminous Supernovae from the Zwicky Transient Facility}

\correspondingauthor{Lin Yan}
\email{lyan@caltech.edu}

\author[0000-0003-1710-9339]{Lin Yan}
\affil{The Caltech Optical Observatories, California Institute of Technology, Pasadena, CA 91125, USA}
\author[0000-0001-8472-1996]{D.~A.~Perley}
\affil{Astrophysics Research Institute, Liverpool John Moores University, 146 Brownlow Hill, Liverpool L3 5RF, UK}
\author{S.~Schulze}
\affil{Department of Particle Physics and Astrophysics, Weizmann Institute of Science, Rehovot 76100, Israel}
\author[0000-0001-9454-4639]{R.~Lunnan}
\affil{The Oskar Klein Centre,  Department of Astronomy, Stockholm University, AlbaNova, SE-106 91 Stockholm, Sweden}
\author[0000-0003-1546-6615]{J.~Sollerman}
\affil{The Oskar Klein Centre, Department of Astronomy, Stockholm University, AlbaNova, SE-106 91 Stockholm, Sweden}
\author{K.~De}
\affil{Division of Physics, Mathematics and Astronomy, California Institute of Technology, Pasadena, CA 91125, USA}
\author[0000-0001-5175-4652]{Z.~H.~Chen}
\affil{Physics Department and Tsinghua Center for Astrophysics (THCA), Tsinghua University, Beijing, 100084, China}
\author{C.~Fremling}
\affil{Division of Physics, Mathematics and Astronomy, California Institute of Technology, Pasadena, CA 91125, USA}
\author{A.~Gal-Yam}
\affil{Department of Particle Physics and Astrophysics, Weizmann Institute of Science, Rehovot 76100, Israel}
\author{K.~Taggart}
\affil{Astrophysics Research Institute, Liverpool John Moores University, 146 Brownlow Hill, Liverpool L3 5RF, UK}
\author[0000-0002-1066-6098]{T.-W.~Chen}
\affil{The Oskar Klein Centre, Department of Astronomy, Stockholm University, AlbaNova, SE-106 91 Stockholm, Sweden}
\author{I.~Andreoni}
\affil{Division of Physics, Mathematics and Astronomy, California Institute of Technology, Pasadena, CA 91125, USA}
\author[0000-0001-8018-5348]{E.~C.~Bellm}
\affiliation{DIRAC Institute, Department of Astronomy, University of Washington, 3910 15th Avenue NE, Seattle, WA 98195, USA} 
\author{V.~Cunningham}
\affil{Department of Astronomy, University of Maryland, College Park, MD 20742, USA}
\author{R.~Dekany}
\affil{The Caltech Optical Observatories, California Institute of Technology, Pasadena, CA 91125, USA}
\author[0000-0001-5060-8733]{D.~A.~Duev}
\affiliation{Division of Physics, Mathematics, and Astronomy, California Institute of Technology, Pasadena, CA 91125, USA}
\author{C.~Fransson}
\affil{The Oskar Klein Centre, Department of Astronomy, Stockholm University, AlbaNova, SE-106 91 Stockholm, Sweden}
\author[0000-0003-2451-5482]{R.~R.~Laher}
\affiliation{IPAC, California Institute of Technology, 1200 E. California
Blvd, Pasadena, CA 91125, USA}
\author{M.~Hankins}
\affil{Division of Physics, Mathematics and Astronomy, California Institute of Technology, Pasadena, CA 91125, USA}
\author[0000-0002-9017-3567]{A.~Y.~Q.~Ho}
\affil{Division of Physics, Mathematics and Astronomy, California Institute of Technology, Pasadena, CA 91125, USA}
\author[0000-0001-5754-4007]{J.~E.~Jencson}
\affiliation{Steward Observatory, University of Arizona, 933 North Cherry Avenue, Tucson, AZ 85721-0065, USA}
\author{S.~Kaye}
\affil{The Caltech Optical Observatories, California Institute of Technology, Pasadena, CA 91125, USA}
\author{S.~R.~Kulkarni}
\affil{Division of Physics, Mathematics and Astronomy, California Institute of Technology, Pasadena, CA 91125, USA}
\author{M.~M.~Kasliwal}
\affil{Division of Physics, Mathematics and Astronomy, California Institute of Technology, Pasadena, CA 91125, USA}
\author[0000-0001-8205-2506]{V.~Z.~Golkhou}
\affiliation{DIRAC Institute, Department of Astronomy, University of Washington, 3910 15th Avenue NE, Seattle, WA 98195, USA} 
\affiliation{The eScience Institute, University of Washington, Seattle, WA 98195, USA}
\author{M.~Graham}
\affil{Division of Physics, Mathematics and Astronomy, California Institute of Technology, Pasadena, CA 91125, USA}
\author[0000-0002-8532-9395]{F.~J.~Masci}
\affiliation{IPAC, California Institute of Technology, 1200 E. California Blvd, Pasadena, CA 91125, USA}
\author[0000-0001-9515-478X]{A.~A.~Miller}
\affiliation{Center for Interdisciplinary Exploration and Research in Astrophysics and Department of Physics and Astronomy, Northwestern University, 1800 Sherman Ave, Evanston, IL 60201, USA}
\affiliation{The Adler Planetarium, Chicago, IL 60605, USA}
\author{J.~D.~Neill}
\affil{Division of Physics, Mathematics and Astronomy, California Institute of Technology, Pasadena, CA 91125, USA}
\author{E.~Ofek}
\affil{Department of Particle Physics and Astrophysics, Weizmann Institute of Science, Rehovot 76100, Israel}
\author{M.~Porter}
\affil{The Caltech Optical Observatories, California Institute of Technology, Pasadena, CA 91125, USA}
\author[0000-0001-7016-1692]{P.~Mr\'oz}
\affil{Division of Physics, Mathematics, and Astronomy, California Institute of Technology, Pasadena, CA 91125, USA}
\author{D.~Reiley}
\affil{The Caltech Optical Observatories, California Institute of Technology, Pasadena, CA 91125, USA}
\author{R.~Riddle}
\affil{The Caltech Optical Observatories, California Institute of Technology, Pasadena, CA 91125, USA}
\author{M.~Rigault}
\affil{Universit\'e Clermont Auvergne, CNRS/IN2P3, Laboratoire de Physique
    de Clermont, F-63000 Clermont-Ferrand, France.}
\author[0000-0001-7648-4142]{B.~Rusholme}
\affiliation{IPAC, California Institute of Technology, 1200 E. California
Blvd, Pasadena, CA 91125, USA}
\author[0000-0003-4401-0430]{D.~L.~Shupe}
\affiliation{IPAC, California Institute of Technology, 1200 E. California
Blvd, Pasadena, CA 91125, USA}
\author[0000-0001-6753-1488]{M.~T.~Soumagnac}
\affiliation{Lawrence Berkeley National Laboratory, 1 Cyclotron Road, Berkeley, CA 94720, USA}
\affiliation{Department of Particle Physics and Astrophysics, Weizmann Institute of Science, Rehovot 76100, Israel}
\author{R.~Smith}
\affil{The Caltech Optical Observatories, California Institute of Technology, Pasadena, CA 91125, USA}
\author{L.~Tartaglia}
\affil{The Oskar Klein Centre, Department of Astronomy, Stockholm University, AlbaNova, SE-106 91 Stockholm, Sweden}
\author[0000-0001-6747-8509]{Y.~Yao}
\affil{Division of Physics, Mathematics and Astronomy, California Institute of Technology, Pasadena, CA 91125, USA}
\author{O.~Yaron}
\affil{Department of Particle Physics and Astrophysics, Weizmann Institute of Science, Rehovot 76100, Israel}


\begin{abstract}
Helium is expected to be present in the massive ejecta of some hydrogen-poor superluminous supernovae (SLSN-I). However, until now only one event has been identified with He features in its photospheric spectra (PTF10hgi, \citealt{Quimby2018}).  We present the discovery of a new He-rich SLSN-I, ZTF19aawfbtg (SN2019hge) at $z=0.0866$.  This event has more than 10 optical spectra at phases from $-41$ to $+103$\,days relative to the peak, most of which match well with that of PTF10hgi. Confirmation comes from a near-IR spectrum taken at $+34$\,days, revealing He\,I features with P-Cygni profiles at 1.083 and 2.058~$\mu$m. Using the optical spectra of PTF10hgi and SN2019hge as templates, we examined 70 other SLSNe-I discovered by ZTF in the first two years of operation and found five additional SLSNe-I with distinct He-features. The excitation of He\,I atoms in normal core collapse supernovae requires non-thermal radiation, as proposed by previous studies. 
These He-rich events cannot be explained by the traditional $^{56}$Ni mixing model because of their blue spectra, high peak luminosities and long rise timescales. 
Magnetar models offer a possible solution since 
pulsar winds naturally generate high energy particles, potential sources of non-thermal excitation. 
An alternative model is interaction between the ejecta and a dense H-poor CSM, which may be supported by observed undulations in the light curves. These six SLSNe-Ib have relatively low-peak luminosities (rest-frame $M_g = -20.06\pm0.16$).

\end{abstract}


\section{Introduction}
Stars more massive than $\sim 8\,M_\odot$ end their lives as core-collapse supernovae (CC SNe \citep{Jones2013}. The observational properties of CC SNe are quite diverse, with much of this variety owing to variations in the mass loss history of the progenitor stars. Hydrogen-rich CC SNe (Type II) result from massive stars that retain their hydrogen envelopes up until the time of explosion, and show hydrogen features in their spectra.  Hydrogen-poor CC SNe (Type I) originate from massive stars that have lost most or all of their H envelopes prior to explosion. These
include SNe IIb (some H \&\ He), SNe Ib (no H, some He), and SNe Ic (no H, no He) \citep{filippenko1997}.

While it is not necessarily a surprise that helium is present within the envelopes of hydrogen-poor SN progenitors, the conditions necessary to produce observable helium features in SN spectra are not always easy to produce, and detections of helium in SNe Ib/IIb have motivated many theoretical  calculations. For example, the \ion{He}{1}\,$\lambda5876$ transition is between energy levels of $2p$ $^3P^0$ and $3d$ $^3D$. To reach these excited states from the ground state ($1s^2$) requires at least 20.9\,eV of energy, corresponding to a temperature of $\sim2.4\times10^5$\,K. As noted by, {\it e.g.} \citet{Dessart2012} and \citet{Lucy1991}, $\gamma$-ray photons from $^{56}$Ni decay can produce high-energy electrons via Compton scattering. These non-thermal electrons are a source of heat, critical for powering the light curve, as well as a source of non-thermal excitation and ionization for the production of optical He\,I lines observed in SNe\,Ib/IIb. This requires the transport of $^{56}$Ni from the inner ejecta to the outer layer where helium material is located. Thus the concept of the $^{56}$Ni ``mixing'' is important for SNe\,Ib/IIb \citep{Dessart2012}.

Since the discovery of the first superluminous supernova more than fifteen years ago \citep[SLSN;][]{Quimby2007}, more than 200 such events have been identified.  A large fraction of low-$z$ events are from two wide-area transient surveys: the Palomar Transient Factory \citep[PTF;][]{Law2009} and the Zwicky Transient Facility \citep[ZTF;][]{Graham2019, Bellm2019}. SLSNe are $10 - 100$ times more luminous than normal CC SNe and often maintain these luminosities for months, requiring additional power beyond what can be provided by the decay of radioactive material \citep{quimby2011,Kasen2010, Yan2015, woosley2017}. As with normal CC SNe, SLSNe are broadly categorized into two classes based on their spectrum around maximum light: hydrogen poor (SLSN-I) and hydrogen rich (SLSN-II) \citep{Gal-Yam2012}.

Although hydrogen-poor SLSNe-I normally do not show hydrogen features in their spectra, a small fraction ($\sim 10-30$\,\%) show intermediate-width H$\alpha$ emission (several 1000\,km\,s$^{-1}$) at $>100$\,days post-peak, presumably produced by ejecta interaction with H-rich circumstellar material (CSM) \citep{Yan2015,Yan2017a}.  These cases suggest that the loss of the progenitor's hydrogen envelope must have occurred shortly before the explosions. The implication is that some helium is likely in the ejecta because there is not enough time to completely strip-off  all of the helium. However, SLSN-I spectral observations have not revealed the presence of helium, with one exception. \citet{Quimby2018} carried out a comprehensive spectral analysis of the entire PTF SLSN-I sample, and identified one event, PTF10hgi, with a clear detection of \ion{He}{1}\,$\lambda\lambda5876,6678,7065,7281$ at $\sim7000$\,km\,s$^{-1}$.  This event also shows probable H$\alpha$ with a broad P-Cygni profile with a velocity of 10000\,km\,s$^{-1}$ at $+47$\,days, suggesting that the event could be classified as a SLSN-IIb. The presence of H$\alpha$ is further supported by a blended emission feature of [O\,I]\,$\lambda6300$ and H$\alpha$ in the spectrum at $+314$\,days.
(SN2012il was proposed to have a He\,I\,1.08$\mu$m feature by \cite{inserra2013}, however \citet{Quimby2018} pointed out that this feature is unlikely to be He\,I, but may be broad nebular H emission.)

ZTF has been operating for over two years, and we have been systematically monitoring the data stream to search for and spectroscopically confirm new SLSNe, resulting in over 100 new SLSN discoveries to date (\citealt{Lunnan2019}; Perley et al. in prep; Chen et al. in prep) and a large accompanying spectral dataset. 
Here we examine this dataset, aiming to address if there are other He-rich SLSN-I similar to PTF10hgi. If such a population of events exists, further study of these events may shed light on the question of how helium atoms are excited and ionized, with important implications for the physical nature of SLSNe-I.

This paper is organized as follows.  \S\ref{sec:ztf19aawopt} and \S\ref{sec:nirspec} present the optical and near-IR spectra for SN2019hge (ZTF19aawfbtg). \S\ref{sec:Ibsample} discusses the sample of SLSN-Ib/IIb events. The paper concludes (\S\ref{sec:disc}) with a discussion of the key results and their implications.
Throughout the paper, all magnitudes are in the AB system unless explicitly noted otherwise. We adopt the $\Lambda$CDM cosmology with H$_0=67.4$\,km\,s$^{-1}$\,Mpc$^{-1}$, $\Omega_M=0.315$ and $\Omega_\Lambda=0.685$ \citep{Planck2018}. 

\begin{figure*}
\centering
\includegraphics[width=19cm]{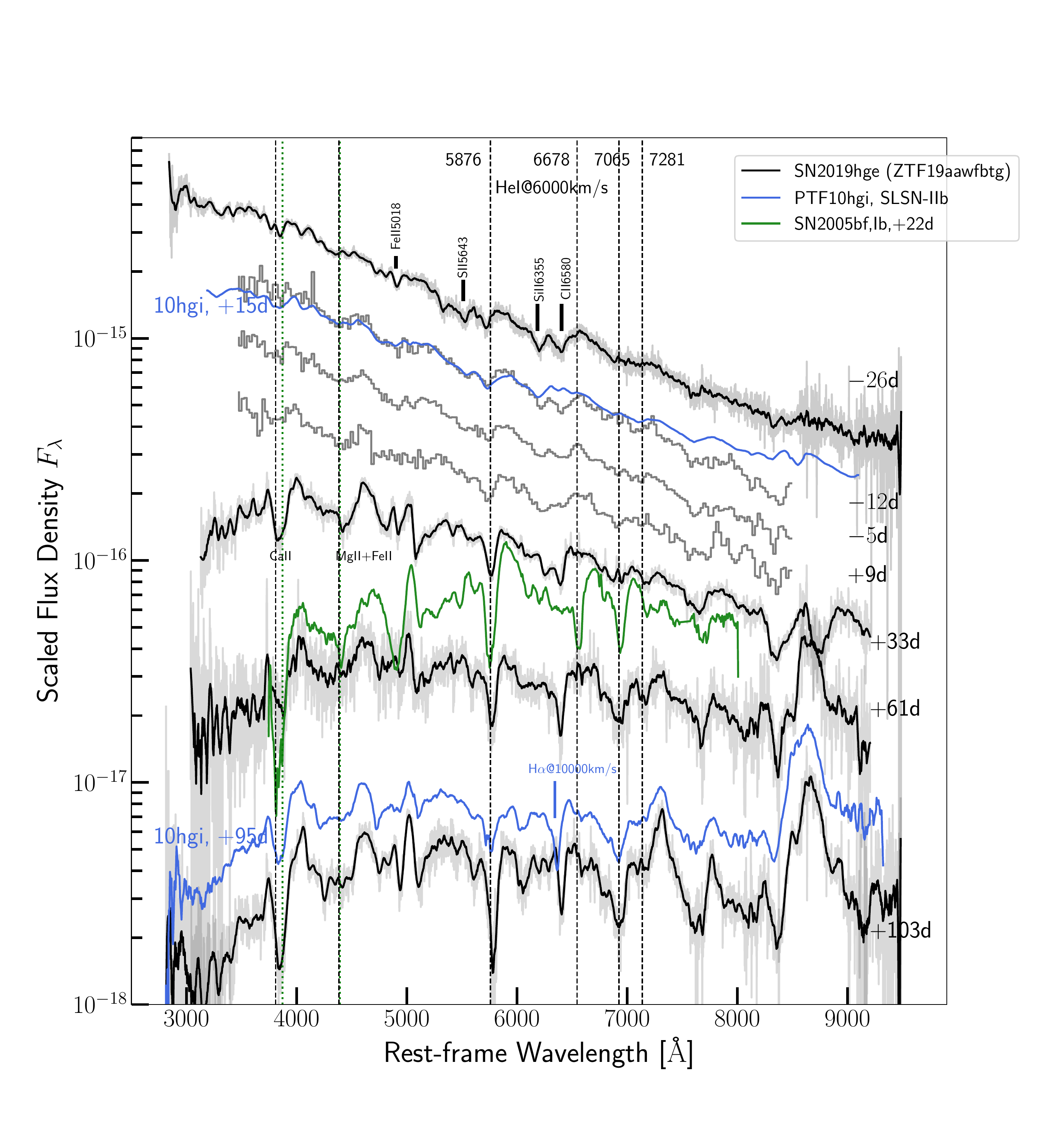}
\caption{Spectra of SN2019hge (ZTF19aawfbtg) around peak and at late phases. Spectra of PTF10hgi (blue) and SN\,2005bf (green) are also shown for comparison. The spectra shown in black are smoothed using the Savitzky Golay filtering algorithm and overlaid onto the spectra in gray, which are the original data. The PTF10hgi and SN\,2005bf spectra are also smoothed with the same method.
The vertical dashed lines mark the locations of \ion{He}{1}\,$\lambda\lambda5876,6678,7065,7281$ at a velocity of 7000\,km\,s$^{-1}$. \label{fig:latespec}}
\end{figure*}

\section{Classification of SN2019hge (ZTF19aawfbgt) as a SLSN-Ib}
\label{sec:ztf19aaw}
\subsection{Optical Spectra}
\label{sec:ztf19aawopt}
Optical spectra of SN2019hge were obtained at $-41$ to $+103$\,days from the $g$-band peak using the low resolution ($R\sim100$) Spectral Energy Distribution Machine \citep[SEDM;][]{Blagorodnova2018} and the Double Beam Spectrograph \citep[DBSP;][]{Oke1983} on the Palomar 60\,inch and 200\,inch telescope respectively, as well as the Low Resolution Imaging Spectrometer \citep[LRIS;][]{Oke1995} on the Keck I telescope.  An observing log for all optical spectra of ZTF19aawfbtg presented here is in Table~\ref{tab:obslog}. Spectral reductions were performed using the SEDM pipeline and the {\it pyraf-dbsp} and {\it LPipe} packages  \citep{Rignault2019,Bellm2016,perley2019}.
Narrow H$\alpha$ and [O\,II]\,$\lambda$3727 emission lines from the host galaxy establish the redshift as $z=0.0866$. The supernova is at the edge of a face-on spiral galaxy with an estimated stellar mass of $\rm log_{10}(M_{*}/M\odot) = 9.51^{+0.15}_{-0.29}$ from the SED fitting of broad band photometry provided by SDSS, PS1, 2MASS and unWISE. This mass is larger than that of a typical SLSN-I host \citep{Perley2016,lunnan2014,Leloudas2015,Schulze2018}.


We applied {\it Superfit} \citep{Howell2013} with the enhanced SLSN spectral templates from \citet{Quimby2018} to our data. We find that for the spectra at phases later than $-26$\,days, all five of the top matches returned by {\it Superfit} are with PTF10hgi.
{\it Superfit}, like all spectral template matching software, can produce ambiguous classifications when the supernova features are weak. Thus, we also manually analysed these spectra to and establish individual spectral line identifications.
Figure~\ref{fig:latespec} compares the spectra of SN2019hge with those of PTF10hgi and the low luminosity Type Ib SN\,2005bf \citep{Folatelli2006}.  There are many similarities between these three events, in particular the four He\,I features at 5876, 6678, 7065 and 7281\,\AA\ blue-shifted with a velocity of $\sim7000$\,km\,s$^{-1}$. We do not consider \ion{He}{1}\,$\lambda\lambda3886, 4472$ as the primary classification lines because of strong contamination from  \ion{Ca}{2}\,$\lambda\lambda3934, 3969$ (H+K lines) and \ion{Mg}{2}\,$\lambda4481$. Of the four He features in Figure~\ref{fig:latespec}, \ion{He}{1}\,$\lambda5876$ is strong with a well defined absorption profile; however, it could have some contamination from \ion{Na}{1}\,$\lambda5890$. The other three features are weak, but \ion{He}{1}\,$\lambda7065$ is well isolated. Therefore, the key to identify He-rich events is the presence of multiple He features, as demonstrated in the classification of normal SNe\,Ib/IIb \citep{Modjaz2014}.

PTF10hgi has He\,I and H$\alpha$ velocities of 7000, and 10,000\,km\,s$^{-1}$ respectively, making it a SLSN-IIb \citep{Quimby2018}. In SN2019hge, hydrogen features are not as clear. In Figure~\ref{fig:latespec}, we marked the two features blueward of 
\ion{He}{1}\,$\lambda6678$ as \ion{C}{2}\,$\lambda6580$ 
and \ion{Si}{2}\,$\lambda6355$. If these were H$\alpha$, the velocity would be either too low ($\sim7800$\,km\,s$^{-1}$) or too high ($\sim17,000$\,km\,s$^{-1}$). We thus conclude that SN2019hge is a H-poor and He-rich event. 
This classification was also noted 
by \citet{Prentice2019} reporting SN\,2019hge as Type IIb based on their optical spectra.

\subsection{Near-IR spectral confirmation of helium in SN2019hge}
\label{sec:nirspec}
\begin{figure*}[t!]
\centering
\includegraphics[width=18cm]{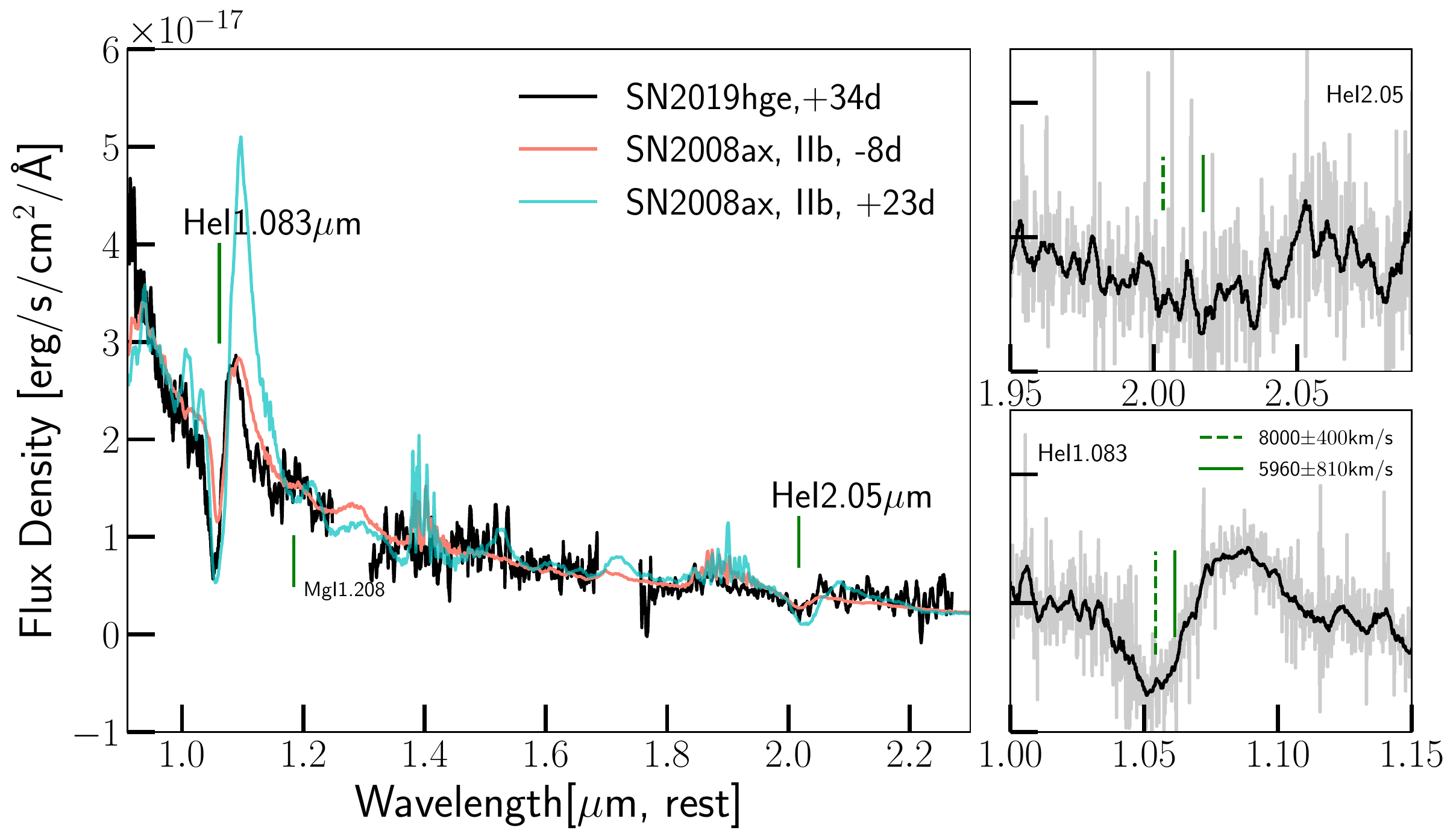}
\caption{The Near-IR spectrum of SN2019hge taken at $+34$ rest-frame days after the $g$-band peak date. The two panels on the right show the zoomed-in view of the two He\,I features. The gray spectra are original data and the spectra in black have been smoothed using the Savitzky–Golay filtering algorithm. The vertical dashed and solid green lines mark two different blueshift velocities: 5960\,km\,s$^{-1}$ measured from the optical spectrum at $+33$\,days, and 8000\,km\,s$^{-1}$ which matches the minimum of the He\,I\,1.083$\mu$m. The pink and cyan lines are the spectra of type IIb SN\,2008ax (no smoothing) at $-8$ and $+23$\,days from the $B$-band maximum light \citep{Taubenberger2011}.  The SN\,2008ax spectra are linearly shifted to match the velocity of 6000\,km\,s$^{-1}$ and no stretching was applied. \label{fig:nirspec}}
\end{figure*}

SN2019hge was observed at $+34$\,days with the Near-Infrared Echellette Spectrometer \citep[NIRES;][]{Wilson2004}, a prism cross-dispersed near-infrared spectrograph mounted on the Keck 2 telescope. This spectrograph has a resolution of $\sim2700$ and simultaneously covers the YJHK bands with a wavelength range of $0.8 - 2.4\,\mu$m. The spectra were reduced with {\it Spextool} and {\it Xtellcor} written by \citet{Cushing2004, Vacca2003}. The data is shown in Figure~\ref{fig:nirspec}, where  
the prominent feature at $1.08\,\mu$m is likely \ion{He}{1}\,1.08303\,$\mu$m with P-Cygni profile. This is confirmed by the presence of \ion{He}{1}\,2.05813\,$\mu$m, a relatively isolated line.  
For comparison, we also plot the near-IR spectra of SN\,2008ax, a SN\,IIb at $-8$ and $+23$\,days from the $B$-band peak \citep{Taubenberger2011}. 
In Figure~\ref{fig:nirspec}, we shift the spectra of SN\,2008ax to a velocity of 6000\,km\,s$^{-1}$.  The match between SN2019hge and SN2008ax is very clear, including weak \ion{Mg}{1}\,1.1828,1.2083\,$\mu$m doublet. The relative strength of the 1.08 and 2.05$\mu$m features for SN2019hge is similar to that of SN2008ax at $-8$\,days.

Assuming the minimum near 1.08$\mu$m is due to the P-Cygni profile of \ion{He}{1}\,1.08303\,$\mu$m, the implied velocity is 8000\,km\,s$^{-1}$, as marked in green dashed vertical lines in the two panels in Figure~\ref{fig:nirspec}. Similarly, the \ion{He}{1}\,2.05813\,$\mu$m P-Cygni profile gives a velocity of 6000\,km\,s$^{-1}$, consistent with that derived from \ion{He}{1}\,$\lambda5876$ at a similar phase of $+33$\,day. It is possible that near-IR He I lines could have higher velocities than the optical He I lines because of their lower excitation. The lower velocity of He I\,$\lambda2.058\,\mu$m could be due to its weaker and noisier feature. An alternative explanation is that \ion{He}{1}\,$1.083\,\mu$m could be blended with \ion{C}{1}\,$1.0691\,\mu$m at 6000\,km\,s$^{-1}$, which would broaden and shift the minimum to shorter wavelengths. C\,I features are also present in the optical spectra at similar phases. The strong $1.08\,\mu$m features in normal SNe\,Ib/Ic could have contributions from He I, C I, Mg II and Si I \citep{Hunter2009,Taubenberger2011}. However, unlike SNe\,Ib, normal SNe\,Ic do not have He I $2.058\,\mu$m \citep{Taubenberger2011}.

If hydrogen is present in the ejecta, Pa$\gamma$ at 1.0941\,$\mu$m would imply a velocity of 11,000\,km\,s$^{-1}$.
Although this velocity is higher as expected, the presence of Pa$\gamma$ is not corroborated by H$\alpha$ and H$\beta$ in the optical spectra. In summary, the optical and near-IR spectra suggest that SN2019hge is a He-rich event.

\section{A sample of He-rich SLSN-I from ZTF}
\label{sec:Ibsample}
\begin{figure*}
    \includegraphics[width=18cm]{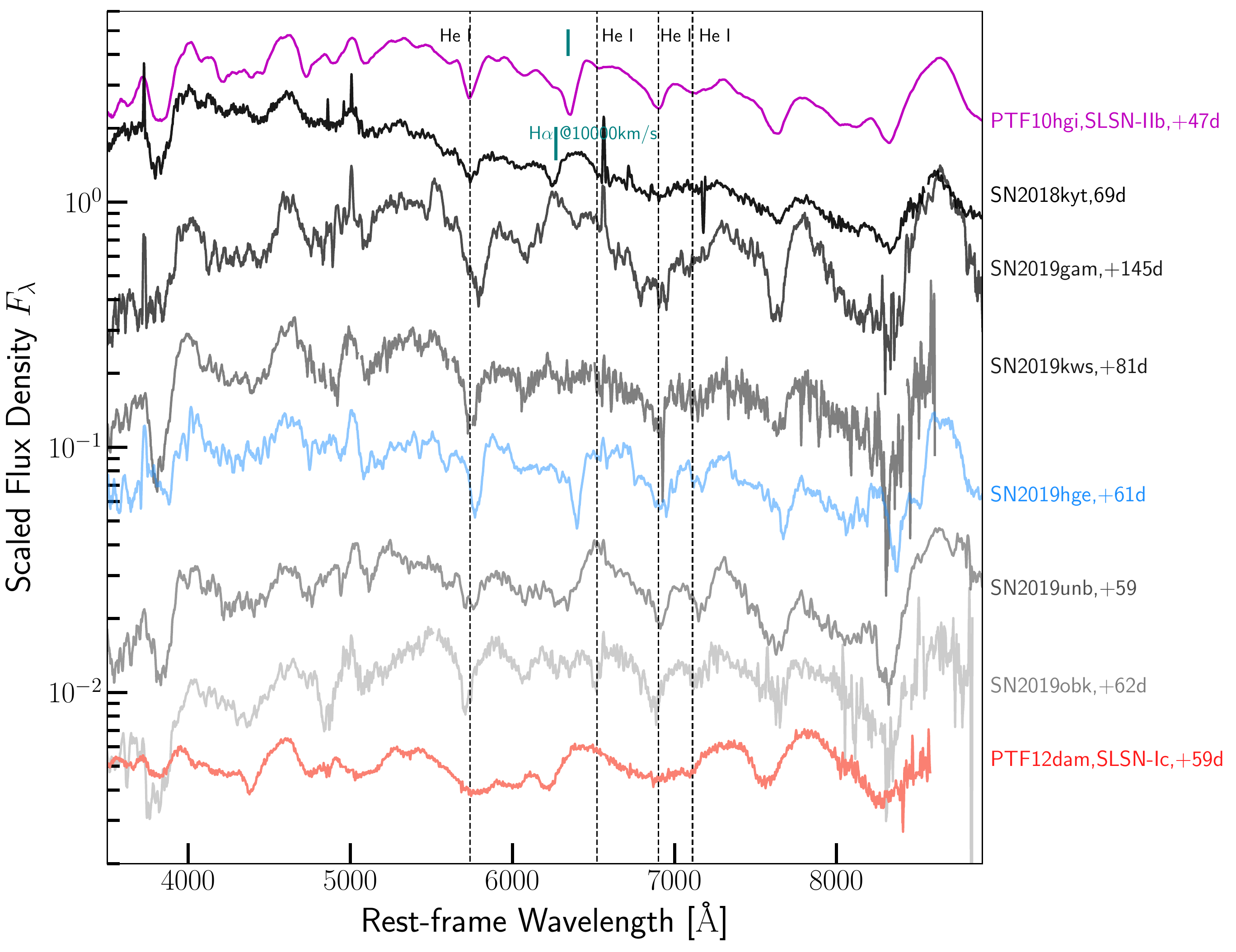}
    \caption{The post-peak spectra for a set of ZTF SLSNe-I are displayed in comparison with that of SN2019hge: PTF10hgi )a SLSN-IIb) and PTF12dam (a SLSN-I). }
    \label{fig:allspec}
\end{figure*}

Assuming that optical spectra of PTF10hgi and SN2019hge can be used as templates to classify SLSN-Ib/IIb, we search for additional sources among the ZTF SLSNe-I discovered during the first two years of the survey. Figure~\ref{fig:allspec} displays the spectra of six SLSNe-I which are matched well with that of PTF10hgi, including SN2019hge.  For these six ZTF events, their post-peak spectra are all matched with that of PTF10hgi among the top 5 ranks when we run {\it Superfit}.  For each of the six events, at least 3 matches from {\it Superfit} are PTF10hgi at various phases, while the remaining 1 or 2 matches are other SLSN-I templates, such as PTF10nmn or PTF11rks. For example, for ZTF19acgjpgh, 4 out of 5 matches are PTF10hgi.  All 5 matches are PTF10hgi for ZTF19aawfbtg.

Figure~\ref{fig:allspec} compares these six events with PTF12dam \citep{Quimby2018}, an archetypal SLSN-I. Near $\lambda \sim 5876$\,\AA\ (He I), PTF12dam does not show a distinctive absorption feature, but a smooth and wide trough suggesting a blend of many lines. This is in contrast with the strong, narrow, and well defined absorption features at 5876\,\AA\ detected in the six ZTF events. More importantly, three other He\,I lines at 6678, 7065 and 7281\,\AA\ are also present, albeit with various strengths. Considering that He-rich material in SN2019hge is confirmed by its near-IR spectrum, we conclude that the other five events are good SLSN-Ib/IIb candidates. In Figure~\ref{fig:allspec}, we note that SN2018kyt (ZTF18acyxnyw) may have H$\alpha$ at $\sim+10,000$\,km\,s$^{-1}$, a He-rich event with a potentially small amount of H, similar to PTF10hgi.

\begin{figure}
    \centering
    \includegraphics[width=8cm]{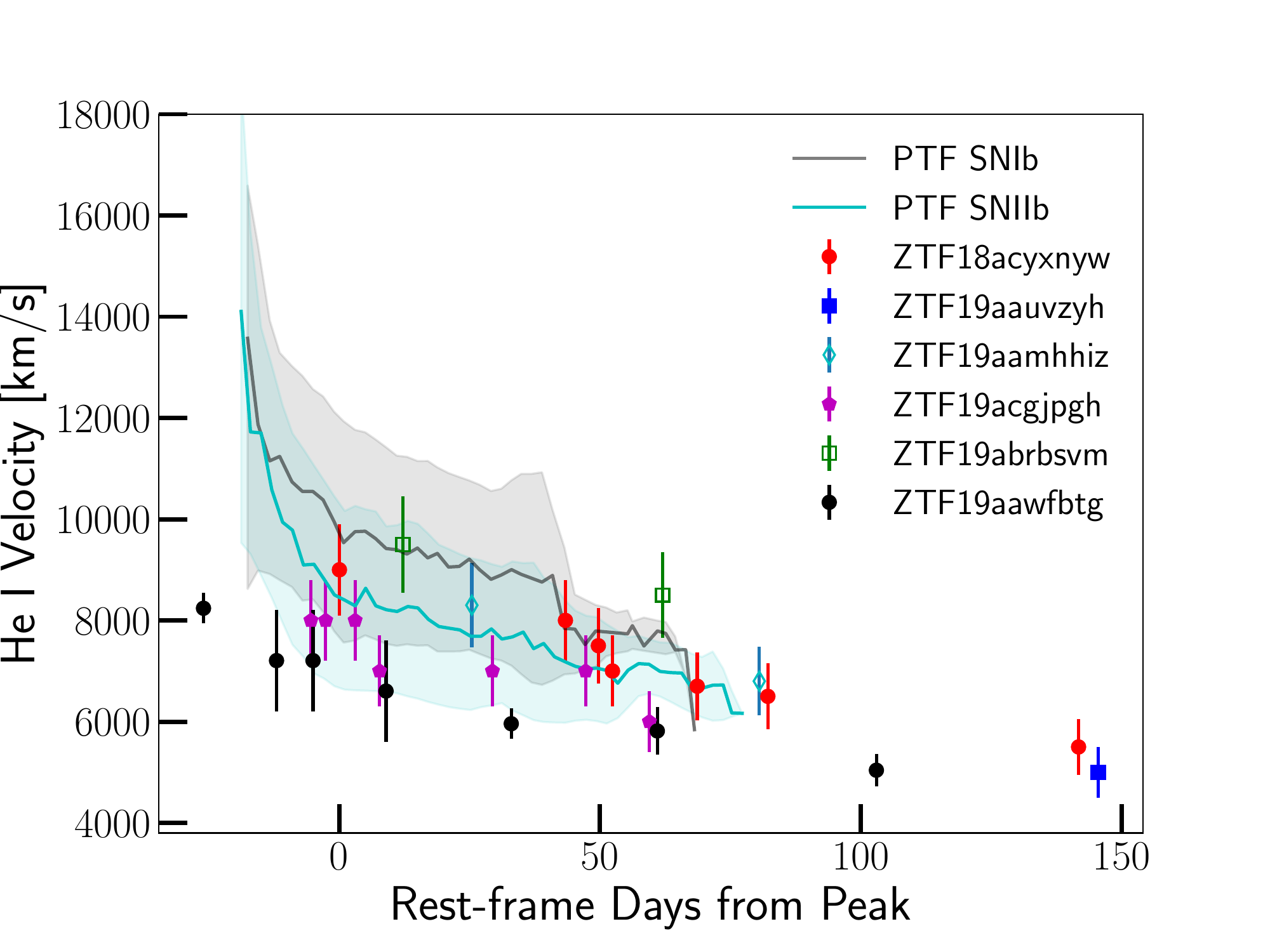}
    \caption{The outflowing velocity of He\,I $\lambda$5876 
    as a function of time measured for the six SLSNe-Ib/IIb.}
    \label{fig:vel}
\end{figure}

If the feature at 5876\,\AA\ is indeed blue-shifted He\,I, we can measure the outflowing velocity $V$ as well as the velocity dispersion by fitting a Gaussian profile to each feature. Figure~\ref{fig:vel} shows the He\,I velocities as a function of time.  The velocities slow down roughly linearly with time, decreasing by 3000\,km\,s$^{-1}$ over 125\,days. For comparison, Figure~\ref{fig:vel} also illustrates that the He I velocities in normal SNe\,Ib and IIb can drop from 14,000\,km\,s$^{-1}$ to 7000 - 6000\,km\,s$^{-1}$ in less than 90\,days, a much more rapid evolution than what is seen in our He-rich SLSN-I sample. Finally, the He\,I line velocity dispersion, {\it i.e.} the line width, is $\sim2000$\,km\,s$^{-1}$ and remains roughly constant over the same period of time. This is much less than the bulk velocity, explaining the narrow appearance of \ion{He}{1}\,$\lambda5876$ line. Finally, \citet{Liu2016} found that that for normal SNe\,Ib, their \ion{He}{1}\,$\lambda6676$ and $\lambda7065$ lines have similar strength. This is not the case for our events, where the 6676 lines are on average weaker. This again suggests a different excitation mechanism from normal SNe\,Ib.  

\begin{figure}
    \centering
    \includegraphics[width=8cm]{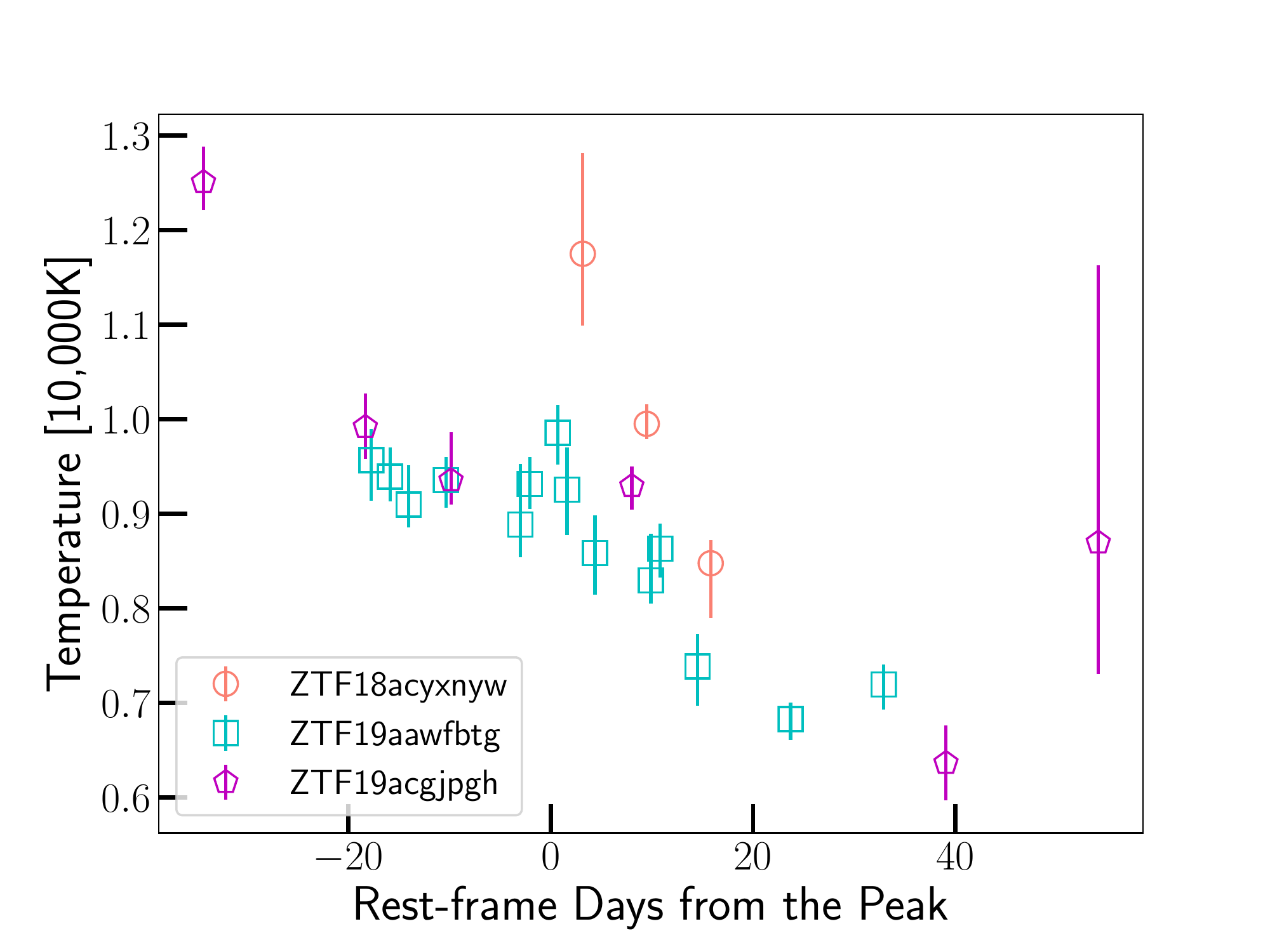}
    \caption{The modified blackbody temperatures measured from broad band photometry including one or more UV-bands. }
    \label{fig:temp}
\end{figure}

In addition, four of these six He-rich events have {\it Swift} UV photometry, and the blackbody temperatures are measured from fitting the broad band UV + optical spectral energy distributions with a modified blackbody function of $B_\lambda^{'} = (\frac{\lambda}{\lambda_0})^\beta \times B_\lambda(T_{BB})$, where $\lambda_0=3000$\,\AA, and $\beta$ and $T_{BB}$ are derived from the fit (details in Chen et al. in prep). The rationale for this function is to take into account various degrees of SED suppression at far to near-UV due to metal line blanketing, as seen in the {\it HST} far-UV spectra of SLSN-I \citep{Yan2017b, Yan2018}. Figure~\ref{fig:temp} shows $T_{BB}$ vs. phases, illustrating that at pre-peak early phases, $T_{BB}$ can reach as high as 12,000\,K, commonly seen among SLSN-I \citep{quimby2011,inserra2013,Nicholl2017,Chen+Nicholl2017}. The He\,I features appear at phases when $T_{BB}$ is lower temperatures, around $\sim8000 - 7000$\,K.

\begin{figure*}
    \centering
    \includegraphics[width=18cm]{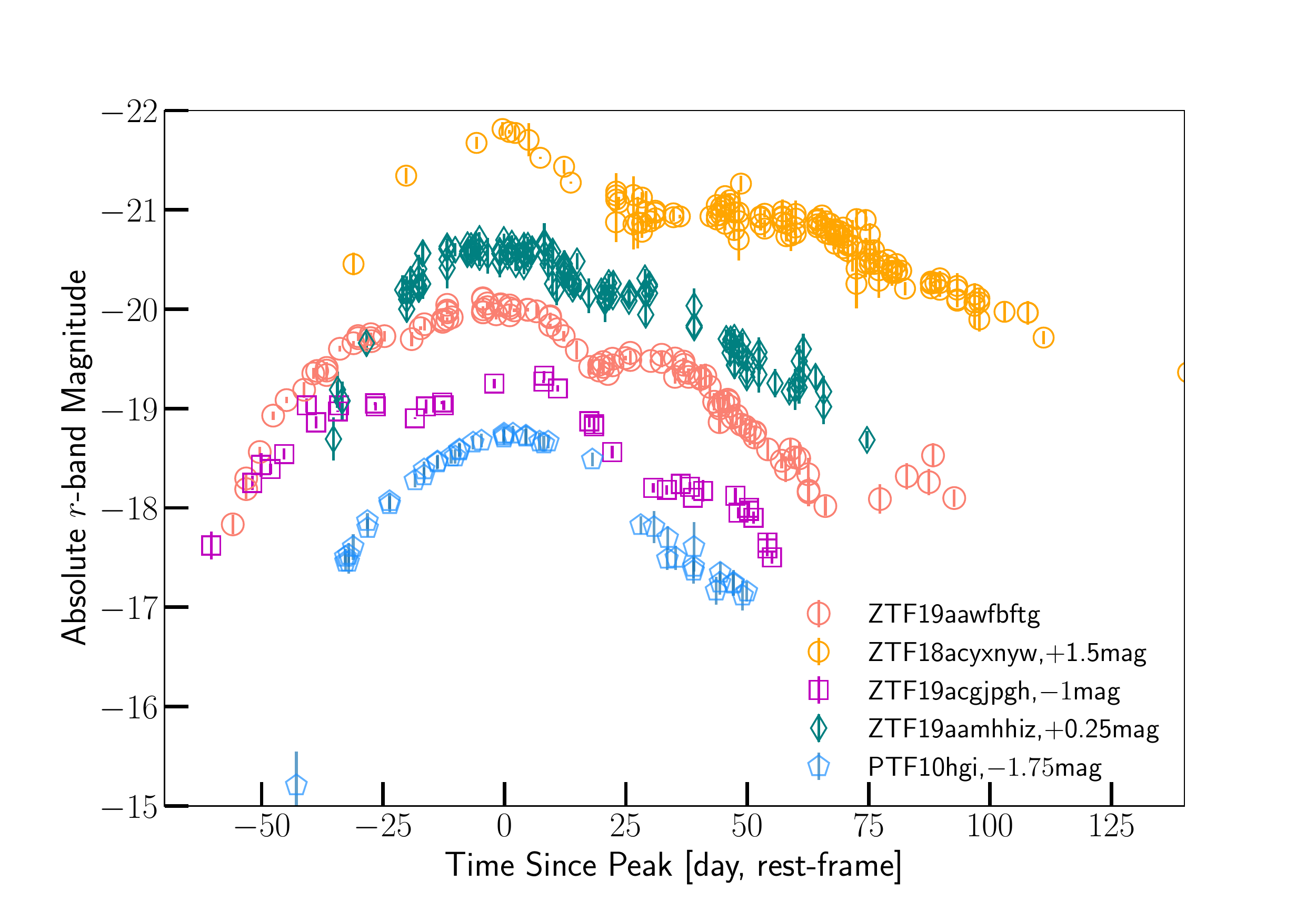}
    \caption{The $r$-band LCs of the five SLSNe-Ib whose LCs have good phase coverage. }
    \label{fig:lcs}
\end{figure*}

Two outstanding features of this sample are shown by the light curves (LC) of the five events with good phase coverage from this sample, as in Figure~\ref{fig:lcs}. Here the LCs were constructed using the ZTF forced photometry utility \footnote{http://web.ipac.caltech.edu/staff/fmasci/ztf/forcedphot.pdf} and additional code from \citet{Yao2019} based on the reference subtracted images from the ZTF image reduction pipeline \citep{Masci2019}. The first notable feature is the prominent LC undulations in four of the six ZTF events (the other two have poor LC coverage). The physical interpretation of this behavior is not clear and there are multiple potential explanations, such as interaction or change in opacity. The second notable feature is the relatively low peak luminosity: all events are in the range of $-19.8$ to $-20.2$\,absolute magnitude, for a sample of slow evolving SLSNe-I with rising time scales of $35 - 68$\,days.  
Table~\ref{tab:meta} lists the name, coordinate, redshift ($z$) and $g$-band absolute peak magnitude (M$_g$). Extinction and K-corrections have been applied. This is in contrast to a much wider range of luminosity, $-20$ to $-23$\,mag, normally occupied by SLSNe-I. Finally, we note that the host galaxies of these seven He-rich events include both low mass dwarf galaxies and massive ones ($10^9.5$), but within the range expected for SLSNe-I \citep{lunnan2014,Perley2016,Schulze2018}.

\section{Discussion and conclusions}
\label{sec:disc}
The main result of this paper is the discovery of a sample of He-rich SLSN-Ib/IIb, all of which are  luminous ($\sim -20$) and slow evolving with very large ejecta masses. 
Based on these six ZTF events, the percentage of He-rich events among ZTF SLSNe-I is $\sim9$\%. This is just an observed fraction of SLSN-I.  Detailed calculations of the volumetric rates including incompleteness are outside the scope of this paper and will be discussed in a separate publication. 

Helium is expected to be present in the ejecta of some massive stars if the stripping of the outer envelopes is incomplete. The detection of late-time, broad H$\alpha$ emission in several PTF SLSNe-I \citep{Yan2015,Yan2017a} has provided the first evidence that some massive stars lost their H-envelope shortly ($<$ a few decades) before the supernova explosion, and never had enough time to also lose their helium layers. The fraction of SLSNe-I with late-time H$\alpha$ emission is roughly 10-30\%\ \citep{Yan2017a}. 

Our result has important implications for both the mass loss history of progenitor stars as well as the power mechanism of SLSNe-I. 
Because the He\,I excitation energy is high, non-thermal excitation is required to produce observable helium features. For normal SNe\,Ib/IIb, non-thermal excitation is achieved by ``mixing'' which transports $^{56}$Ni from the interior to the outer parts of ejecta \citep{Lucy1991,Dessart2012}. This solution is unlikely to work for these SLSN-Ib because having $^{56}$Ni as the sole power source would require extremely large $^{56}$Ni masses, given the observed LCs with high peak luminosities and long rise timescales. Furthermore, Fe-group elements from $^{56}$Ni decay will produce significant line blanketing in the UV/optical, leading to red spectra, in contrast to our observations (Figure~\ref{fig:latespec}).
This argument also rules out the pair-instability supernova model \citep[PISN;][]{woosley2007}. 
For SLSNe-I, central engine models such as a magnetar or black hole fall-back offer a natural source of high energy particles from either pulsar wind or outflow from accretion disk. These outflow winds will interact with the inner regions of ejecta, producing shocks and X-rays. There are two spectral formation models for SLSN-I. One by \citet{Mazzali2016} assumes X-rays from shocks driven by magnetar wind as a source of non-thermal excitation. The other by \citet{Dessart2019} adopts a sustained magnetar energy injection profile (spatial). Both models are able to explain the hallmark O\,II absorption features at 3500 - 4500\AA\   \citep{quimby2011} and even produce He\,I\,$\lambda5876$\,\AA. However, both models do not deal with the dynamic processes of magnetar wind propagating through the ejecta. So the details of how magnetar energy is transported and thermalized to the outer layers are not understood. Furthermore, X-ray emission has not been detected from SLSNe-I (except one peculiar case) and current limits \citep{margutti2018} do not support the idea that shock interaction is an important energy source \citep{Mazzali2016}. The recent 3D magnetar simulation by \citet{Chen2020} has examined the dynamical processes, suggesting that Rayleigh-Taylor instability can quickly drive Ni and Fe to the outer layers. However, their work does not include simulated spectra. It is likely that Fe-group elements will make the spectra too red. Finally, ejecta-CSM interaction is a possible solution. The observed LC undulations from our sample may support this scenario. However, the LC undulations could also be related to the ejecta clumpiness and changes in opacity.

An important question is why only a small fraction of SLSN-I have He\,I features. Is the fraction of progenitors retaining some helium at the time of the explosion intrinsically small? If true, this would suggest very efficient mass loss mechanisms. If massive stars are mostly in binaries, stripping by the secondary accretion is certainly more efficient than wind mass loss. An alternative scenario is that most SLSN-I progenitors have helium, but only a small fraction show He features in their spectra, either due to lack of excitation energy, or other peculiar physical conditions. Future modeling work may address this question.
  
Another spectral feature worth noting is that the early phase spectra of these events are blue continuum dominated without obvious O\,II absorption features (most notably, at $-41$\,d for SN2019hge and $-28$\,d for PTF10hgi).  
This cannot be explained by low temperatures: the early phase temperatures are clearly very high (Figure~\ref{fig:temp}). 
Spectral smearing due to high velocity is also unlikely to provide a complete explanation because spectra taken post-peak all show narrow features with relatively low velocities (Figure~\ref{fig:latespec} \&\ ~\ref{fig:allspec}). Like He\,I features, the formation of O\,II absorption requires not only sufficient excitation energy to produce ionized O$^+$, but also is affected by other factors, such as density and ionization parameter. However, the existing models have not provided clear physical explanations for why sometimes He\,I and O\,II absorption features are formed and sometimes they are not.

\acknowledgements

Based on observations obtained with the Samuel Oschin Telescope 48-inch and the 60-inch Telescope at the Palomar Observatory as part of the Zwicky Transient Facility project. ZTF is supported by the National Science Foundation under Grant No. AST-1440341 and a collaboration including Caltech, IPAC, the Weizmann Institute for Science, the Oskar Klein Center at Stockholm University, the University of Maryland, the University of Washington, Deutsches Elektronen-Synchrotron and Humboldt University, Los Alamos National Laboratories, the TANGO Consortium of Taiwan, the University of Wisconsin at Milwaukee, and Lawrence Berkeley National Laboratories. Operations are conducted by COO, IPAC, and UW. The Liverpool Telescope is operated on the island of La Palma by Liverpool John Moores University in the Spanish Observatorio del Roque de los Muchachos of the Instituto de Astrofisica de Canarias with financial support from the UK Science and Technology Facilities Council. SED Machine is based upon work supported by the National Science Foundation under Grant No. 1106171.
The ZTF forced-photometry service was funded under the Heising-Simons Foundation grant \#12540303 (PI: Graham).
This work uses the GROWTH Followup Marshal \citep{Mansi2019} and was supported by the GROWTH project funded by the National Science Foundation under Grant No 1545949. A.A.M.~is funded by the Large Synoptic Survey Telescope Corporation, the
Brinson Foundation, and the Moore Foundation in support of the LSSTC Data
Science Fellowship Program; he also receives support as a CIERA Fellow by the
CIERA Postdoctoral Fellowship Program (Center for Interdisciplinary
Exploration and Research in Astrophysics, Northwestern University).
R.L. is supported by a Marie Sk\l{}odowska-Curie Individual Fellowship within the Horizon 2020 European Union (EU) Framework Programme for Research and Innovation (H2020-MSCA-IF-2017-794467).

\bibliography{references}

\begin{thebibliography}{}
\expandafter\ifx\csname natexlab\endcsname\relax\def\natexlab#1{#1}\fi
\providecommand{\url}[1]{\href{#1}{#1}}
\providecommand{\dodoi}[1]{}
\providecommand{\doarXiv}[1]{\href{https://arxiv.org/abs/#1}{\nolinkurl{https://arxiv.org/abs/#1}}}

\bibitem[{{Bellm} \& {Sesar}(2016)}]{Bellm2016}
{Bellm}, E.~C., \& {Sesar}, B. 2016, {pyraf-dbsp: Reduction pipeline for the
  Palomar Double Beam Spectrograph}

\bibitem[{{Bellm} {et~al.}(2019){Bellm}, {Kulkarni}, {Graham}, {Dekany},
  {Smith}, {Riddle}, {Masci}, {Helou}, {Prince}, {Adams}, {Barbarino},
  {Barlow}, {Bauer}, {Beck}, {Belicki}, {Biswas}, {Blagorodnova}, {Bodewits},
  {Bolin}, {Brinnel}, {Brooke}, {Bue}, {Bulla}, {Burruss}, {Cenko}, {Chang},
  {Connolly}, {Coughlin}, {Cromer}, {Cunningham}, {De}, {Delacroix}, {Desai},
  {Duev}, {Eadie}, {Farnham}, {Feeney}, {Feindt}, {Flynn}, {Franckowiak},
  {Frederick}, {Fremling}, {Gal-Yam}, {Gezari}, {Giomi}, {Goldstein},
  {Golkhou}, {Goobar}, {Groom}, {Hacopians}, {Hale}, {Henning}, {Ho}, {Hover},
  {Howell}, {Hung}, {Huppenkothen}, {Imel}, {Ip}, {Ivezi{\'c}}, {Jackson},
  {Jones}, {Juric}, {Kasliwal}, {Kaspi}, {Kaye}, {Kelley}, {Kowalski},
  {Kramer}, {Kupfer}, {Landry}, {Laher}, {Lee}, {Lin}, {Lin}, {Lunnan},
  {Giomi}, {Mahabal}, {Mao}, {Miller}, {Monkewitz}, {Murphy}, {Ngeow},
  {Nordin}, {Nugent}, {Ofek}, {Patterson}, {Penprase}, {Porter}, {Rauch},
  {Rebbapragada}, {Reiley}, {Rigault}, {Rodriguez}, {van Roestel}, {Rusholme},
  {van Santen}, {Schulze}, {Shupe}, {Singer}, {Soumagnac}, {Stein}, {Surace},
  {Sollerman}, {Szkody}, {Taddia}, {Terek}, {Van Sistine}, {van Velzen},
  {Vestrand}, {Walters}, {Ward}, {Ye}, {Yu}, {Yan}, \& {Zolkower}}]{Bellm2019}
{Bellm}, E.~C., {Kulkarni}, S.~R., {Graham}, M.~J., {et~al.} 2019,
  \href{http://dx.doi.org/10.1088/1538-3873/aaecbe}{\color{magenta}\pasp},
  \href{https://ui.adsabs.harvard.edu/abs/2019PASP..131a8002B}{\color{blue}131},
  \href{https://ui.adsabs.harvard.edu/abs/2019PASP..131a8002B}{\color{blue}018002}

\bibitem[{{Blagorodnova} {et~al.}(2018){Blagorodnova}, {Neill}, {Walters},
  {Kulkarni}, {Fremling}, {Ben-Ami}, {Dekany}, {Fucik}, {Konidaris}, {Nash},
  {Ngeow}, {Ofek}, {O' Sullivan}, {Quimby}, {Ritter}, \&
  {Vyhmeister}}]{Blagorodnova2018}
{Blagorodnova}, N., {Neill}, J.~D., {Walters}, R., {et~al.} 2018,
  \href{http://dx.doi.org/10.1088/1538-3873/aaa53f}{\color{magenta}\pasp},
  \href{https://ui.adsabs.harvard.edu/abs/2018PASP..130c5003B}{\color{blue}130},
  \href{https://ui.adsabs.harvard.edu/abs/2018PASP..130c5003B}{\color{blue}035003}

\bibitem[{{Chen} {et~al.}(2020){Chen}, {Woosley}, \& {Whalen}}]{Chen2020}
{Chen}, K.-J., {Woosley}, S.~E., \& {Whalen}, D.~J. 2020,
  \href{http://dx.doi.org/10.3847/1538-4357/ab7db0}{\color{magenta}\apj},
  \href{https://ui.adsabs.harvard.edu/abs/2020ApJ...893...99C}{\color{blue}893},
  \href{https://ui.adsabs.harvard.edu/abs/2020ApJ...893...99C}{\color{blue}99}

\bibitem[{{Chen} {et~al.}(2017){Chen}, {Nicholl}, {Smartt}, {Mazzali}, {Yates},
  {Moriya}, {Inserra}, {Langer}, {Kr{\"u}hler}, {Pan}, {Kotak}, {Galbany},
  {Schady}, {Wiseman}, {Greiner}, {Schulze}, {Man}, {Jerkstrand }, {Smith},
  {Dennefeld}, {Baltay}, {Bolmer}, {Kankare}, {Knust}, {Maguire}, {Rabinowitz},
  {Rostami}, {Sullivan}, \& {Young}}]{Chen+Nicholl2017}
{Chen}, T.~W., {Nicholl}, M., {Smartt}, S.~J., {et~al.} 2017,
  \href{http://dx.doi.org/10.1051/0004-6361/201630163}{\color{magenta}\aap},
  \href{https://ui.adsabs.harvard.edu/abs/2017A&A...602A...9C}{\color{blue}602},
  \href{https://ui.adsabs.harvard.edu/abs/2017A&A...602A...9C}{\color{blue}A9}

\bibitem[{{Cushing} {et~al.}(2004){Cushing}, {Vacca}, \&
  {Rayner}}]{Cushing2004}
{Cushing}, M.~C., {Vacca}, W.~D., \& {Rayner}, J.~T. 2004,
  \href{http://dx.doi.org/10.1086/382907}{\color{magenta}\pasp},
  \href{https://ui.adsabs.harvard.edu/abs/2004PASP..116..362C}{\color{blue}116},
  \href{https://ui.adsabs.harvard.edu/abs/2004PASP..116..362C}{\color{blue}362}

\bibitem[{{De Cia} {et~al.}(2018){De Cia}, {Gal-Yam}, {Rubin}, {Leloudas},
  {Vreeswijk}, {Perley}, {Quimby}, {Yan}, {Sullivan}, {Fl{\"o}rs}, {Sollerman},
  {Bersier}, {Cenko}, {Gal-Yam}, {Maguire}, {Ofek}, {Prentice}, {Schulze},
  {Spyromilio}, {Valenti}, {Arcavi}, {Corsi}, {Howell}, {Mazzali}, {Kasliwal},
  {Taddia}, \& {Yaron}}]{DeCia2018}
{De Cia}, A., {Gal-Yam}, A., {Rubin}, A., {et~al.} 2018,
  \href{http://dx.doi.org/10.3847/1538-4357/aab9b6}{\color{magenta}\apj},
  \href{https://ui.adsabs.harvard.edu/abs/2018ApJ...860..100D}{\color{blue}860},
  \href{https://ui.adsabs.harvard.edu/abs/2018ApJ...860..100D}{\color{blue}100}

\bibitem[{{Dessart}(2019)}]{Dessart2019}
{Dessart}, L. 2019,
  \href{http://dx.doi.org/10.1051/0004-6361/201834535}{\color{magenta}\aap},
  \href{https://ui.adsabs.harvard.edu/abs/2019A&A...621A.141D}{\color{blue}621},
  \href{https://ui.adsabs.harvard.edu/abs/2019A&A...621A.141D}{\color{blue}A141}

\bibitem[{{Dessart} {et~al.}(2012){Dessart}, {Hillier}, {Waldman}, {Livne}, \&
  {Blondin}}]{Dessart2012}
{Dessart}, L., {Hillier}, D.~J., {Waldman}, R., {et~al.} 2012,
  \href{http://dx.doi.org/10.1111/j.1745-3933.2012.01329.x}{\color{magenta}\mnras},
  \href{http://adsabs.harvard.edu/abs/2012MNRAS.426L..76D}{\color{blue}426},
  \href{http://adsabs.harvard.edu/abs/2012MNRAS.426L..76D}{\color{blue}L76}

\bibitem[{Filippenko(1997)}]{filippenko1997}
Filippenko, A. 1997, \araa, 35, 309

\bibitem[{{Folatelli} {et~al.}(2006){Folatelli}, {Contreras}, {Phillips},
  {Woosley}, {Blinnikov}, {Morrell}, {Suntzeff}, {Lee}, {Hamuy},
  {Gonz{\'a}lez}, {Krzeminski}, {Roth}, {Li}, {Filippenko}, {Foley},
  {Freedman}, {Madore}, {Persson}, {Murphy}, {Boissier}, {Galaz},
  {Gonz{\'a}lez}, {McCarthy}, {McWilliam}, \& {Pych}}]{Folatelli2006}
{Folatelli}, G., {Contreras}, C., {Phillips}, M.~M., {et~al.} 2006,
  \href{http://dx.doi.org/10.1086/500531}{\color{magenta}\apj},
  \href{https://ui.adsabs.harvard.edu/abs/2006ApJ...641.1039F}{\color{blue}641},
  \href{https://ui.adsabs.harvard.edu/abs/2006ApJ...641.1039F}{\color{blue}1039}

\bibitem[{{Gal-Yam}(2012)}]{Gal-Yam2012}
{Gal-Yam}, A. 2012,
  \href{http://dx.doi.org/10.1126/science.1203601}{\color{magenta}Science},
  \href{https://ui.adsabs.harvard.edu/abs/2012Sci...337..927G}{\color{blue}337},
  \href{https://ui.adsabs.harvard.edu/abs/2012Sci...337..927G}{\color{blue}927}

\bibitem[{{Graham} {et~al.}(2019){Graham}, {Kulkarni}, {Bellm}, {Adams},
  {Barbarino}, {Blagorodnova}, {Bodewits}, {Bolin}, {Brady}, {Cenko}, {Chang},
  {Coughlin}, {De}, {Eadie}, {Farnham}, {Feindt}, {Franckowiak}, {Fremling},
  {Gezari}, {Ghosh}, {Goldstein}, {Golkhou}, {Goobar}, {Ho}, {Huppenkothen},
  {Ivezi{\'c}}, {Jones}, {Juric}, {Kaplan}, {Kasliwal}, {Kelley}, {Kupfer},
  {Lee}, {Lin}, {Lunnan}, {Mahabal}, {Miller}, {Ngeow}, {Nugent}, {Ofek},
  {Prince}, {Rauch}, {van Roestel}, {Schulze}, {Singer}, {Sollerman}, {Taddia},
  {Yan}, {Ye}, {Yu}, {Barlow}, {Bauer}, {Beck}, {Belicki}, {Biswas}, {Brinnel},
  {Brooke}, {Bue}, {Bulla}, {Burruss}, {Connolly}, {Cromer}, {Cunningham},
  {Dekany}, {Delacroix}, {Desai}, {Duev}, {Feeney}, {Flynn}, {Frederick},
  {Gal-Yam}, {Giomi}, {Groom}, {Hacopians}, {Hale}, {Helou}, {Henning},
  {Hover}, {Hillenbrand}, {Howell}, {Hung}, {Imel}, {Ip}, {Jackson}, {Kaspi},
  {Kaye}, {Kowalski}, {Kramer}, {Kuhn}, {Landry}, {Laher}, {Mao}, {Masci},
  {Monkewitz}, {Murphy}, {Nordin}, {Patterson}, {Penprase}, {Porter},
  {Rebbapragada}, {Reiley}, {Riddle}, {Rigault}, {Rodriguez}, {Rusholme}, {van
  Santen}, {Shupe}, {Smith}, {Soumagnac}, {Stein}, {Surace}, {Szkody}, {Terek},
  {Van Sistine}, {van Velzen}, {Vestrand}, {Walters}, {Ward}, {Zhang}, \&
  {Zolkower}}]{Graham2019}
{Graham}, M.~J., {Kulkarni}, S.~R., {Bellm}, E.~C., {et~al.} 2019,
  \href{http://dx.doi.org/10.1088/1538-3873/ab006c}{\color{magenta}\pasp},
  \href{https://ui.adsabs.harvard.edu/abs/2019PASP..131g8001G}{\color{blue}131},
  \href{https://ui.adsabs.harvard.edu/abs/2019PASP..131g8001G}{\color{blue}078001}

\bibitem[{{Howell} {et~al.}(2013){Howell}, {Kasen}, {Lidman}, {Sullivan},
  {Conley}, {Astier}, {Balland}, {Carlberg}, {Fouchez}, {Guy}, {Hardin},
  {Pain}, {Palanque-Delabrouille}, {Perrett}, {Pritchet}, {Regnault}, {Rich},
  \& {Ruhlmann-Kleider}}]{Howell2013}
{Howell}, D.~A., {Kasen}, D., {Lidman}, C., {et~al.} 2013,
  \href{http://dx.doi.org/10.1088/0004-637X/779/2/98}{\color{magenta}\apj},
  \href{http://adsabs.harvard.edu/abs/2013ApJ...779...98H}{\color{blue}779},
  \href{http://adsabs.harvard.edu/abs/2013ApJ...779...98H}{\color{blue}98}

\bibitem[{{Hunter} {et~al.}(2009){Hunter}, {Valenti}, {Kotak}, {Meikle},
  {Taubenberger}, {Pastorello}, {Benetti}, {Stanishev}, {Smartt}, {Trundle},
  {Arkharov}, {Bufano}, {Cappellaro}, {Di Carlo}, {Dolci}, {Elias-Rosa},
  {Frandsen}, {Fynbo}, {Hopp}, {Larionov}, {Laursen}, {Mazzali}, {Navasardyan},
  {Ries}, {Riffeser}, {Rizzi}, {Tsvetkov}, {Turatto}, \& {Wilke}}]{Hunter2009}
{Hunter}, D.~J., {Valenti}, S., {Kotak}, R., {et~al.} 2009,
  \href{http://dx.doi.org/10.1051/0004-6361/200912896}{\color{magenta}\aap},
  \href{https://ui.adsabs.harvard.edu/abs/2009A&A...508..371H}{\color{blue}508},
  \href{https://ui.adsabs.harvard.edu/abs/2009A&A...508..371H}{\color{blue}371}

\bibitem[{Inserra {et~al.}(2013)Inserra, Smartt, Jerkstrand, Valenti, Fraser,
  Wright, Smith, Chen, Kotak, Pastorello, {et~al.}}]{inserra2013}
Inserra, C., Smartt, S., Jerkstrand, A., {et~al.} 2013, \apj, 770, 128

\bibitem[{{Jones} {et~al.}(2013){Jones}, {Hirschi}, {Nomoto}, {Fischer},
  {Timmes}, {Herwig}, {Paxton}, {Toki}, {Suzuki}, {Mart{\'\i}nez-Pinedo},
  {Lam}, \& {Bertolli}}]{Jones2013}
{Jones}, S., {Hirschi}, R., {Nomoto}, K., {et~al.} 2013,
  \href{http://dx.doi.org/10.1088/0004-637X/772/2/150}{\color{magenta}\apj},
  \href{https://ui.adsabs.harvard.edu/abs/2013ApJ...772..150J}{\color{blue}772},
  \href{https://ui.adsabs.harvard.edu/abs/2013ApJ...772..150J}{\color{blue}150}

\bibitem[{{Kasen} \& {Bildsten}(2010)}]{Kasen2010}
{Kasen}, D., \& {Bildsten}, L. 2010,
  \href{http://dx.doi.org/10.1088/0004-637X/717/1/245}{\color{magenta}\apj},
  \href{https://ui.adsabs.harvard.edu/abs/2010ApJ...717..245K}{\color{blue}717},
  \href{https://ui.adsabs.harvard.edu/abs/2010ApJ...717..245K}{\color{blue}245}

\bibitem[{{Kasliwal} {et~al.}(2019){Kasliwal}, {Cannella}, {Bagdasaryan},
  {Hung}, {Feindt}, {Singer}, {Coughlin}, {Fremling}, {Walters}, {Duev},
  {Itoh}, \& {Quimby}}]{Mansi2019}
{Kasliwal}, M.~M., {Cannella}, C., {Bagdasaryan}, A., {et~al.} 2019,
  \href{http://dx.doi.org/10.1088/1538-3873/aafbc2}{\color{magenta}\pasp},
  \href{https://ui.adsabs.harvard.edu/abs/2019PASP..131c8003K}{\color{blue}131},
  \href{https://ui.adsabs.harvard.edu/abs/2019PASP..131c8003K}{\color{blue}038003}

\bibitem[{{Law} {et~al.}(2009){Law}, {Kulkarni}, {Dekany}, {Ofek}, {Quimby},
  {Nugent}, {Surace}, {Grillmair}, {Bloom}, {Kasliwal}, {Bildsten}, {Brown},
  {Cenko}, {Ciardi}, {Croner}, {Djorgovski}, {van Eyken}, {Filippenko}, {Fox},
  {Gal-Yam}, {Hale}, {Hamam}, {Helou}, {Henning}, {Howell}, {Jacobsen},
  {Laher}, {Mattingly}, {McKenna}, {Pickles}, {Poznanski}, {Rahmer}, {Rau},
  {Rosing}, {Shara}, {Smith}, {Starr}, {Sullivan}, {Velur}, {Walters}, \&
  {Zolkower}}]{Law2009}
{Law}, N.~M., {Kulkarni}, S.~R., {Dekany}, R.~G., {et~al.} 2009,
  \href{http://dx.doi.org/10.1086/648598}{\color{magenta}\pasp},
  \href{http://adsabs.harvard.edu/abs/2009PASP..121.1395L}{\color{blue}121},
  \href{http://adsabs.harvard.edu/abs/2009PASP..121.1395L}{\color{blue}1395}

\bibitem[{{Leloudas} {et~al.}(2015){Leloudas}, {Schulze}, {Kr{\"u}hler},
  {Gorosabel}, {Christensen}, {Mehner}, {de Ugarte Postigo}, {Amor{\'{\i}}n},
  {Th{\"o}ne}, {Anderson}, {Bauer}, {Gallazzi}, {He{\l}miniak}, {Hjorth},
  {Ibar}, {Malesani}, {Morell}, {Vinko}, \& {Wheeler}}]{Leloudas2015}
{Leloudas}, G., {Schulze}, S., {Kr{\"u}hler}, T., {et~al.} 2015,
  \href{http://dx.doi.org/10.1093/mnras/stv320}{\color{magenta}\mnras},
  \href{http://adsabs.harvard.edu/abs/2015MNRAS.449..917L}{\color{blue}449},
  \href{http://adsabs.harvard.edu/abs/2015MNRAS.449..917L}{\color{blue}917}

\bibitem[{{Liu} {et~al.}(2016){Liu}, {Modjaz}, {Bianco}, \& {Graur}}]{Liu2016}
{Liu}, Y.-Q., {Modjaz}, M., {Bianco}, F.~B., \& {Graur}, O. 2016,
  \href{http://dx.doi.org/10.3847/0004-637X/827/2/90}{\color{magenta}\apj},
  \href{https://ui.adsabs.harvard.edu/abs/2016ApJ...827...90L}{\color{blue}827},
  \href{https://ui.adsabs.harvard.edu/abs/2016ApJ...827...90L}{\color{blue}90}

\bibitem[{{Lucy}(1991)}]{Lucy1991}
{Lucy}, L.~B. 1991,
  \href{http://dx.doi.org/10.1086/170787}{\color{magenta}\apj},
  \href{https://ui.adsabs.harvard.edu/abs/1991ApJ...383..308L}{\color{blue}383},
  \href{https://ui.adsabs.harvard.edu/abs/1991ApJ...383..308L}{\color{blue}308}

\bibitem[{{Lunnan} {et~al.}(2014){Lunnan}, {Chornock}, {Berger}, {Laskar},
  {Fong}, {Rest}, {Sanders}, {Challis}, {Drout}, {Foley}, {Huber}, {Kirshner},
  {Leibler}, {Marion}, {McCrum}, {Milisavljevic}, {Narayan}, {Scolnic},
  {Smartt}, {Smith}, {Soderberg}, {Tonry}, {Burgett}, {Chambers}, {Flewelling},
  {Hodapp}, {Kaiser}, {Magnier}, {Price}, \& {Wainscoat}}]{lunnan2014}
{Lunnan}, R., {Chornock}, R., {Berger}, E., {et~al.} 2014,
  \href{http://dx.doi.org/10.1088/0004-637X/787/2/138}{\color{magenta}\apj},
  \href{http://adsabs.harvard.edu/abs/2014ApJ...787..138L}{\color{blue}787},
  \href{http://adsabs.harvard.edu/abs/2014ApJ...787..138L}{\color{blue}138}

\bibitem[{{Lunnan} {et~al.}(2019){Lunnan}, {Yan}, {Perley}, {Schulze},
  {Taggart}, {Gal-Yam}, {Fremling}, {Soumagnac}, {Ofek}, {Adams}, {Barbarino},
  {Bellm}, {De}, {Fransson}, {Frederick}, {Golkhou}, {Graham}, {Hallakoun},
  {Ho}, {Kasliwal}, {Kaspi}, {Kulkarni}, {Laher}, {Masci}, {Pozo Nunez},
  {Rusholme}, {Quimby}, {Shupe}, {Sollerman}, {Taddia}, {van Roestel}, {Yang},
  \& {Yao}}]{Lunnan2019}
{Lunnan}, R., {Yan}, L., {Perley}, D.~A., {et~al.} 2019,
  \href{https://arxiv.org/abs/1910.02968}{\color{magenta}arXiv},
  \href{https://ui.adsabs.harvard.edu/abs/2019arXiv191002968L}{\color{blue}arXiv:1910.02968}

\bibitem[{{Margutti} {et~al.}(2018){Margutti}, {Chornock}, {Metzger},
  {Coppejans}, {Guidorzi}, {Migliori}, {Milisavljevic}, {Berger}, {Nicholl},
  {Zauderer}, {Lunnan}, {Kamble}, {Drout}, \& {Modjaz}}]{margutti2018}
{Margutti}, R., {Chornock}, R., {Metzger}, B.~D., {et~al.} 2018,
  \href{http://dx.doi.org/10.3847/1538-4357/aad2df}{\color{magenta}\apj},
  \href{https://ui.adsabs.harvard.edu/abs/2018ApJ...864...45M}{\color{blue}864},
  \href{https://ui.adsabs.harvard.edu/abs/2018ApJ...864...45M}{\color{blue}45}

\bibitem[{{Masci} {et~al.}(2019){Masci}, {Laher}, {Rusholme}, {Shupe}, {Groom},
  {Surace}, {Jackson}, {Monkewitz}, {Beck}, {Flynn}, {Terek}, {Landry},
  {Hacopians}, {Desai}, {Howell}, {Brooke}, {Imel}, {Wachter}, {Ye}, {Lin},
  {Cenko}, {Cunningham}, {Rebbapragada}, {Bue}, {Miller}, {Mahabal}, {Bellm},
  {Patterson}, {Juri{\'c}}, {Golkhou}, {Ofek}, {Walters}, {Graham}, {Kasliwal},
  {Dekany}, {Kupfer}, {Burdge}, {Cannella}, {Barlow}, {Van Sistine}, {Giomi},
  {Fremling}, {Blagorodnova}, {Levitan}, {Riddle}, {Smith}, {Helou}, {Prince},
  \& {Kulkarni}}]{Masci2019}
{Masci}, F.~J., {Laher}, R.~R., {Rusholme}, B., {et~al.} 2019,
  \href{http://dx.doi.org/10.1088/1538-3873/aae8ac}{\color{magenta}\pasp},
  \href{https://ui.adsabs.harvard.edu/abs/2019PASP..131a8003M}{\color{blue}131},
  \href{https://ui.adsabs.harvard.edu/abs/2019PASP..131a8003M}{\color{blue}018003}

\bibitem[{{Mazzali} {et~al.}(2016){Mazzali}, {Sullivan}, {Pian}, {Greiner}, \&
  {Kann}}]{Mazzali2016}
{Mazzali}, P.~A., {Sullivan}, M., {Pian}, E., {et~al.} 2016,
  \href{http://dx.doi.org/10.1093/mnras/stw512}{\color{magenta}\mnras},
  \href{https://ui.adsabs.harvard.edu/abs/2016MNRAS.458.3455M}{\color{blue}458},
  \href{https://ui.adsabs.harvard.edu/abs/2016MNRAS.458.3455M}{\color{blue}3455}

\bibitem[{{Modjaz} {et~al.}(2014){Modjaz}, {Blondin}, {Kirshner}, {Matheson},
  {Berlind}, {Bianco}, {Calkins}, {Challis}, {Garnavich}, {Hicken}, {Jha},
  {Liu}, \& {Marion}}]{Modjaz2014}
{Modjaz}, M., {Blondin}, S., {Kirshner}, R.~P., {et~al.} 2014,
  \href{http://dx.doi.org/10.1088/0004-6256/147/5/99}{\color{magenta}\aj},
  \href{https://ui.adsabs.harvard.edu/abs/2014AJ....147...99M}{\color{blue}147},
  \href{https://ui.adsabs.harvard.edu/abs/2014AJ....147...99M}{\color{blue}99}

\bibitem[{{Nicholl} {et~al.}(2017){Nicholl}, {Berger}, {Margutti}, {Blanchard},
  {Guillochon}, {Leja}, \& {Chornock}}]{Nicholl2017}
{Nicholl}, M., {Berger}, E., {Margutti}, R., {et~al.} 2017,
  \href{http://dx.doi.org/10.3847/2041-8213/aa82b1}{\color{magenta}\apjl},
  \href{http://adsabs.harvard.edu/abs/2017ApJ...845L...8N}{\color{blue}845},
  \href{http://adsabs.harvard.edu/abs/2017ApJ...845L...8N}{\color{blue}L8}

\bibitem[{{Oke} \& {Gunn}(1983)}]{Oke1983}
{Oke}, J.~B., \& {Gunn}, J.~E. 1983,
  \href{http://dx.doi.org/10.1086/160817}{\color{magenta}\apj},
  \href{http://adsabs.harvard.edu/abs/1983ApJ...266..713O}{\color{blue}266},
  \href{http://adsabs.harvard.edu/abs/1983ApJ...266..713O}{\color{blue}713}

\bibitem[{{Oke} {et~al.}(1995){Oke}, {Cohen}, {Carr}, {Cromer}, {Dingizian},
  {Harris}, {Labrecque}, {Lucinio}, {Schaal}, {Epps}, \& {Miller}}]{Oke1995}
{Oke}, J.~B., {Cohen}, J.~G., {Carr}, M., {et~al.} 1995,
  \href{http://dx.doi.org/10.1086/133562}{\color{magenta}\pasp},
  \href{http://adsabs.harvard.edu/abs/1995PASP..107..375O}{\color{blue}107},
  \href{http://adsabs.harvard.edu/abs/1995PASP..107..375O}{\color{blue}375}

\bibitem[{{Perley}(2019)}]{perley2019}
{Perley}, D.~A. 2019,
  \href{http://dx.doi.org/10.1088/1538-3873/ab215d}{\color{magenta}\pasp},
  \href{https://ui.adsabs.harvard.edu/abs/2019PASP..131h4503P}{\color{blue}131},
  \href{https://ui.adsabs.harvard.edu/abs/2019PASP..131h4503P}{\color{blue}084503}

\bibitem[{{Perley} {et~al.}(2016){Perley}, {Quimby}, {Yan}, {Vreeswijk}, {De
  Cia}, {Lunnan}, {Gal-Yam}, {Yaron}, {Filippenko}, {Graham}, {Laher}, \&
  {Nugent}}]{Perley2016}
{Perley}, D.~A., {Quimby}, R.~M., {Yan}, L., {et~al.} 2016,
  \href{http://dx.doi.org/10.3847/0004-637X/830/1/13}{\color{magenta}\apj},
  \href{http://adsabs.harvard.edu/abs/2016ApJ...830...13P}{\color{blue}830},
  \href{http://adsabs.harvard.edu/abs/2016ApJ...830...13P}{\color{blue}13}

\bibitem[{{Planck Collaboration} {et~al.}(2018){Planck Collaboration},
  {Aghanim}, {Akrami}, {Ashdown}, {Aumont}, {Baccigalupi}, {Ballardini},
  {Banday}, {Barreiro}, {Bartolo}, {Basak}, {Battye}, {Benabed}, {Bernard},
  {Bersanelli}, {Bielewicz}, {Bock}, {Bond}, {Borrill}, {Bouchet}, {Boulanger},
  {Bucher}, {Burigana}, {Butler}, {Calabrese}, {Cardoso}, {Carron},
  {Challinor}, {Chiang}, {Chluba}, {Colombo}, {Combet}, {Contreras}, {Crill},
  {Cuttaia}, {de Bernardis}, {de Zotti}, {Delabrouille}, {Delouis}, {Di
  Valentino}, {Diego}, {Dor{\'e}}, {Douspis}, {Ducout}, {Dupac}, {Dusini},
  {Efstathiou}, {Elsner}, {En{\ss}lin}, {Eriksen}, {Fantaye}, {Farhang},
  {Fergusson}, {Fernandez-Cobos}, {Finelli}, {Forastieri}, {Frailis},
  {Fraisse}, {Franceschi}, {Frolov}, {Galeotta}, {Galli}, {Ganga},
  {G{\'e}nova-Santos}, {Gerbino}, {Ghosh}, {Gonz{\'a}lez-Nuevo}, {G{\'o}rski},
  {Gratton}, {Gruppuso}, {Gudmundsson}, {Hamann}, {Handley}, {Hansen},
  {Herranz}, {Hildebrandt}, {Hivon}, {Huang}, {Jaffe}, {Jones}, {Karakci},
  {Keih{\"a}nen}, {Keskitalo}, {Kiiveri}, {Kim}, {Kisner}, {Knox},
  {Krachmalnicoff}, {Kunz}, {Kurki-Suonio}, {Lagache}, {Lamarre}, {Lasenby},
  {Lattanzi}, {Lawrence}, {Le Jeune}, {Lemos}, {Lesgourgues}, {Levrier},
  {Lewis}, {Liguori}, {Lilje}, {Lilley}, {Lindholm}, {L{\'o}pez-Caniego},
  {Lubin}, {Ma}, {Mac{\'\i}as-P{\'e}rez}, {Maggio}, {Maino}, {Mandolesi},
  {Mangilli}, {Marcos-Caballero}, {Maris}, {Martin}, {Martinelli},
  {Mart{\'\i}nez-Gonz{\'a}lez}, {Matarrese}, {Mauri}, {McEwen}, {Meinhold},
  {Melchiorri}, {Mennella}, {Migliaccio}, {Millea}, {Mitra},
  {Miville-Desch{\^e}nes}, {Molinari}, {Montier}, {Morgante}, {Moss}, {Natoli},
  {N{\o}rgaard-Nielsen}, {Pagano}, {Paoletti}, {Partridge}, {Patanchon},
  {Peiris}, {Perrotta}, {Pettorino}, {Piacentini}, {Polastri}, {Polenta},
  {Puget}, {Rachen}, {Reinecke}, {Remazeilles}, {Renzi}, {Rocha}, {Rosset},
  {Roudier}, {Rubi{\~n}o-Mart{\'\i}n}, {Ruiz-Granados}, {Salvati}, {Sandri},
  {Savelainen}, {Scott}, {Shellard}, {Sirignano}, {Sirri}, {Spencer},
  {Sunyaev}, {Suur-Uski}, {Tauber}, {Tavagnacco}, {Tenti}, {Toffolatti},
  {Tomasi}, {Trombetti}, {Valenziano}, {Valiviita}, {Van Tent}, {Vibert},
  {Vielva}, {Villa}, {Vittorio}, {Wand elt}, {Wehus}, {White}, {White},
  {Zacchei}, \& {Zonca}}]{Planck2018}
{Planck Collaboration}, {Aghanim}, N., {Akrami}, Y., {et~al.} 2018,
  \href{https://arxiv.org/abs/1807.06209}{\color{magenta}arXiv},
  \href{https://ui.adsabs.harvard.edu/abs/2018arXiv180706209P}{\color{blue}arXiv:1807.06209}

\bibitem[{{Prentice} {et~al.}(2019){Prentice}, {Maguire}, {Skillen}, {Magee},
  \& {Clark}}]{Prentice2019}
{Prentice}, S.~J., {Maguire}, K., {Skillen}, K., {et~al.} 2019, Transient Name
  Server AstroNote,
  \href{https://ui.adsabs.harvard.edu/abs/2019TNSAN..74....1P}{\color{blue}74},
  \href{https://ui.adsabs.harvard.edu/abs/2019TNSAN..74....1P}{\color{blue}1}

\bibitem[{{Quimby} {et~al.}(2007){Quimby}, {Aldering}, {Wheeler},
  {H{\"o}flich}, {Akerlof}, \& {Rykoff}}]{Quimby2007}
{Quimby}, R.~M., {Aldering}, G., {Wheeler}, J.~C., {et~al.} 2007,
  \href{http://dx.doi.org/10.1086/522862}{\color{magenta}\apjl},
  \href{http://cdsads.u-strasbg.fr/abs/2007ApJ...668L..99Q}{\color{blue}668},
  \href{http://cdsads.u-strasbg.fr/abs/2007ApJ...668L..99Q}{\color{blue}L99}

\bibitem[{{Quimby} {et~al.}(2011){Quimby}, {Kulkarni}, {Kasliwal}, {Gal-Yam},
  {Arcavi}, {Sullivan}, {Nugent}, {Thomas}, {Howell}, {Nakar}, {Bildsten},
  {Theissen}, {Law}, {Dekany}, {Rahmer}, {Hale}, {Smith}, {Ofek}, {Zolkower},
  {Velur}, {Walters}, {Henning}, {Bui}, {McKenna}, {Poznanski}, {Cenko}, \&
  {Levitan}}]{quimby2011}
{Quimby}, R.~M., {Kulkarni}, S.~R., {Kasliwal}, M.~M., {et~al.} 2011,
  \href{http://dx.doi.org/10.1038/nature10095}{\color{magenta}\nat},
  \href{https://ui.adsabs.harvard.edu/abs/2011Natur.474..487Q}{\color{blue}474},
  \href{https://ui.adsabs.harvard.edu/abs/2011Natur.474..487Q}{\color{blue}487}

\bibitem[{{Quimby} {et~al.}(2018){Quimby}, {De Cia}, {Gal-Yam}, {Leloudas},
  {Lunnan}, {Perley}, {Vreeswijk}, {Yan}, {Bloom}, {Cenko}, {Cooke}, {Ellis},
  {Filippenko}, {Kasliwal}, {Kleiser}, {Kulkarni}, {Matheson}, {Nugent}, {Pan},
  {Silverman}, {Sternberg}, {Sullivan}, \& {Yaron}}]{Quimby2018}
{Quimby}, R.~M., {De Cia}, A., {Gal-Yam}, A., {et~al.} 2018,
  \href{http://dx.doi.org/10.3847/1538-4357/aaac2f}{\color{magenta}\apj},
  \href{https://ui.adsabs.harvard.edu/abs/2018ApJ...855....2Q}{\color{blue}855},
  \href{https://ui.adsabs.harvard.edu/abs/2018ApJ...855....2Q}{\color{blue}2}

\bibitem[{{Rigault} {et~al.}(2019){Rigault}, {Neill}, {Blagorodnova}, {Dugas},
  {Feeney}, {Walters}, {Brinnel}, {Copin}, {Fremling}, {Nordin}, \&
  {Sollerman}}]{Rignault2019}
{Rigault}, M., {Neill}, J.~D., {Blagorodnova}, N., {et~al.} 2019,
  \href{http://dx.doi.org/10.1051/0004-6361/201935344}{\color{magenta}\aap},
  \href{https://ui.adsabs.harvard.edu/abs/2019A&A...627A.115R}{\color{blue}627},
  \href{https://ui.adsabs.harvard.edu/abs/2019A&A...627A.115R}{\color{blue}A115}

\bibitem[{{Schulze} {et~al.}(2018){Schulze}, {Kr{\"u}hler}, {Leloudas},
  {Gorosabel}, {Mehner}, {Buchner}, {Kim}, {Ibar}, {Amor{\'{\i}}n},
  {Herrero-Illana}, {Anderson}, {Bauer}, {Christensen}, {de Pasquale}, {de
  Ugarte Postigo}, {Gallazzi}, {Hjorth}, {Morrell}, {Malesani}, {Sparre},
  {Stalder}, {Stark}, {Th{\"o}ne}, \& {Wheeler}}]{Schulze2018}
{Schulze}, S., {Kr{\"u}hler}, T., {Leloudas}, G., {et~al.} 2018,
  \href{http://dx.doi.org/10.1093/mnras/stx2352}{\color{magenta}\mnras},
  \href{http://adsabs.harvard.edu/abs/2018MNRAS.473.1258S}{\color{blue}473},
  \href{http://adsabs.harvard.edu/abs/2018MNRAS.473.1258S}{\color{blue}1258}

\bibitem[{{Taubenberger} {et~al.}(2011){Taubenberger}, {Navasardyan}, {Maurer},
  {Zampieri}, {Chugai}, {Benetti}, {Agnoletto}, {Bufano}, {Elias-Rosa},
  {Turatto}, {Patat}, {Cappellaro}, {Mazzali}, {Iijima}, {Valenti},
  {Harutyunyan}, {Claudi}, \& {Dolci}}]{Taubenberger2011}
{Taubenberger}, S., {Navasardyan}, H., {Maurer}, J.~I., {et~al.} 2011,
  \href{http://dx.doi.org/10.1111/j.1365-2966.2011.18287.x}{\color{magenta}\mnras},
  \href{https://ui.adsabs.harvard.edu/abs/2011MNRAS.413.2140T}{\color{blue}413},
  \href{https://ui.adsabs.harvard.edu/abs/2011MNRAS.413.2140T}{\color{blue}2140}

\bibitem[{{Vacca} {et~al.}(2003){Vacca}, {Cushing}, \& {Rayner}}]{Vacca2003}
{Vacca}, W.~D., {Cushing}, M.~C., \& {Rayner}, J.~T. 2003,
  \href{http://dx.doi.org/10.1086/346193}{\color{magenta}\pasp},
  \href{https://ui.adsabs.harvard.edu/abs/2003PASP..115..389V}{\color{blue}115},
  \href{https://ui.adsabs.harvard.edu/abs/2003PASP..115..389V}{\color{blue}389}

\bibitem[{{Wilson} {et~al.}(2004){Wilson}, {Henderson}, {Herter}, {Matthews},
  {Skrutskie}, {Adams}, {Moon}, {Smith}, {Gautier}, {Ressler}, {Soifer}, {Lin},
  {Howard}, {LaMarr}, {Stolberg}, \& {Zink}}]{Wilson2004}
{Wilson}, J.~C., {Henderson}, C.~P., {Herter}, T.~L., {et~al.} 2004,
  \href{http://dx.doi.org/10.1117/12.550925}{\color{magenta}Society of
  Photo-Optical Instrumentation Engineers (SPIE) Conference Series}, Vol.
  \href{https://ui.adsabs.harvard.edu/abs/2004SPIE.5492.1295W}{\color{blue}5492},
  {Mass producing an efficient NIR spectrograph}, ed. A.~F.~M. {Moorwood} \&
  M.~{Iye},
  \href{https://ui.adsabs.harvard.edu/abs/2004SPIE.5492.1295W}{\color{blue}1295}

\bibitem[{{Woosley}(2017)}]{woosley2017}
{Woosley}, S.~E. 2017,
  \href{http://dx.doi.org/10.3847/1538-4357/836/2/244}{\color{magenta}\apj},
  \href{https://ui.adsabs.harvard.edu/abs/2017ApJ...836..244W}{\color{blue}836},
  \href{https://ui.adsabs.harvard.edu/abs/2017ApJ...836..244W}{\color{blue}244}

\bibitem[{{Woosley} {et~al.}(2007){Woosley}, {Blinnikov}, \&
  {Heger}}]{woosley2007}
{Woosley}, S.~E., {Blinnikov}, S., \& {Heger}, A. 2007,
  \href{http://dx.doi.org/10.1038/nature06333}{\color{magenta}\nat},
  \href{https://ui.adsabs.harvard.edu/abs/2007Natur.450..390W}{\color{blue}450},
  \href{https://ui.adsabs.harvard.edu/abs/2007Natur.450..390W}{\color{blue}390}

\bibitem[{{Yan} {et~al.}(2018){Yan}, {Perley}, {De Cia}, {Quimby}, {Lunnan},
  {Rubin}, \& {Brown}}]{Yan2018}
{Yan}, L., {Perley}, D.~A., {De Cia}, A., {et~al.} 2018,
  \href{http://dx.doi.org/10.3847/1538-4357/aabad5}{\color{magenta}\apj},
  \href{http://adsabs.harvard.edu/abs/2018ApJ...858...91Y}{\color{blue}858},
  \href{http://adsabs.harvard.edu/abs/2018ApJ...858...91Y}{\color{blue}91}

\bibitem[{{Yan} {et~al.}(2015){Yan}, {Quimby}, {Ofek}, {Gal-Yam}, {Mazzali},
  {Perley}, {Vreeswijk}, {Leloudas}, {De Cia}, {Masci}, {Cenko}, {Cao},
  {Kulkarni}, {Nugent}, {Rebbapragada}, {Wo{\'z}niak}, \& {Yaron}}]{Yan2015}
{Yan}, L., {Quimby}, R., {Ofek}, E., {et~al.} 2015,
  \href{http://dx.doi.org/10.1088/0004-637X/814/2/108}{\color{magenta}\apj},
  \href{http://adsabs.harvard.edu/abs/2015ApJ...814..108Y}{\color{blue}814},
  \href{http://adsabs.harvard.edu/abs/2015ApJ...814..108Y}{\color{blue}108}

\bibitem[{{Yan} {et~al.}(2017{\natexlab{a}}){Yan}, {Lunnan}, {Perley},
  {Gal-Yam}, {Yaron}, {Roy}, {Quimby}, {Sollerman}, {Fremling}, {Leloudas},
  {Cenko}, {Vreeswijk}, {Graham}, {Howell}, {De Cia}, {Ofek}, {Nugent},
  {Kulkarni}, {Hosseinzadeh}, {Masci}, {McCully}, {Rebbapragada}, \&
  {Wo{\'z}niak}}]{Yan2017a}
{Yan}, L., {Lunnan}, R., {Perley}, D.~A., {et~al.} 2017{\natexlab{a}},
  \href{http://dx.doi.org/10.3847/1538-4357/aa8993}{\color{magenta}\apj},
  \href{https://ui.adsabs.harvard.edu/abs/2017ApJ...848....6Y}{\color{blue}848},
  \href{https://ui.adsabs.harvard.edu/abs/2017ApJ...848....6Y}{\color{blue}6}

\bibitem[{{Yan} {et~al.}(2017{\natexlab{b}}){Yan}, {Quimby}, {Gal-Yam},
  {Brown}, {Blagorodnova}, {Ofek}, {Lunnan}, {Cooke}, {Cenko}, {Jencson}, \&
  {Kasliwal}}]{Yan2017b}
{Yan}, L., {Quimby}, R., {Gal-Yam}, A., {et~al.} 2017{\natexlab{b}},
  \href{http://dx.doi.org/10.3847/1538-4357/aa6b02}{\color{magenta}\apj},
  \href{http://adsabs.harvard.edu/abs/2017ApJ...840...57Y}{\color{blue}840},
  \href{http://adsabs.harvard.edu/abs/2017ApJ...840...57Y}{\color{blue}57}

\bibitem[{{Yao} {et~al.}(2019){Yao}, {Miller}, {Kulkarni}, {Bulla}, {Masci},
  {Goldstein}, {Goobar}, {Nugent}, {Dugas}, {Blagorodnova}, {Neill}, {Rigault},
  {Sollerman}, {Nordin}, {Bellm}, {Cenko}, {De}, {Dhawan}, {Feindt},
  {Fremling}, {Gatkine}, {Graham}, {Graham}, {Ho}, {Hung}, {Kasliwal},
  {Kupfer}, {Laher}, {Perley}, {Rusholme}, {Shupe}, {Soumagnac}, {Taggart},
  {Walters}, \& {Yan}}]{Yao2019}
{Yao}, Y., {Miller}, A.~A., {Kulkarni}, S.~R., {et~al.} 2019,
  \href{http://dx.doi.org/10.3847/1538-4357/ab4cf5}{\color{magenta}\apj},
  \href{https://ui.adsabs.harvard.edu/abs/2019ApJ...886..152Y}{\color{blue}886},
  \href{https://ui.adsabs.harvard.edu/abs/2019ApJ...886..152Y}{\color{blue}152}

\end{thebibliography}
\begin{deluxetable}{cccccc}
\tablecaption{Optical spectra for ZTF19aawfbtg$^a$ \label{tab:obslog}}
\tablewidth{0pt}
\tabletypesize{\scriptsize}
\tablehead{
\colhead{Obs.date} &
\colhead{Phase$^b$} &
\colhead{tel/instr} &
\colhead{Exp.Time} &
\colhead{$\Delta \lambda$} &
\colhead{$R^c$} \\
\colhead{} &
\colhead{days} &
\colhead{} &
\colhead{seconds} &
\colhead{\AA} &
\colhead{} 
}
\startdata
2019-07-02 & -26 & Keck/LRIS  & 300    & 3100 - 10000 &  1000  \\
2019-07-18 & -12 & P60/SEDM   & 1800   & 3900 - 9000  &  100   \\
2019-07-26 & -5  & P60/SEDM   & 1800   & 3900 - 9000  &  100   \\
2019-08-10 & +9d & P60/SEDM   & 2250   & 3900 - 9000  &  100   \\
2019-09-06 & +33 & P200/DBSP  & 900    & 3300 - 9800  &  956   \\
2019-10-06 & +61 & P200/DBSP  & 900    & 3300 - 9800  &  956   \\
2019-11-21 & +103 & Keck/LRIS & 500    & 3900 - 9000  & 1000 \\
\enddata

\tablecomments{{\it \bf a}: All spectral observations were taken using parallactic angles. {\it \bf b}: Phase is the rest-frame days from the $g$-band peak in Julian date of 2458695.72. {\it \bf c}: Spectral resolution $R$ is estimated around 7000\AA. }

\end{deluxetable}

\begin{deluxetable}{cccccc}
\tablecaption{A Sample of ZTF SLSN-Ib/IIb \label{tab:meta}}
\tablewidth{0pt}
\tabletypesize{\scriptsize}
\tablehead{
\colhead{Name} &
\colhead{IAU Name} &
\colhead{RA} &
\colhead{DEC} &
\colhead{$z$} &
\colhead{M$_g$} \\ 
}
\startdata
PTF10hgi & ... & 16:37:47.04 & +06:12:32.3 & 0.0985  & -20.2$^a$ \\
ZTF18acyxnyw & SN2018kyt & 12:27:56.23 & +56:23:35.6 & 0.108 & -20.18 \\
ZTF19aamhhiz & SN2019kws & 14:15:04.46 & +50:39:06.8 & 0.1977 & -20.17 \\
ZTF19aauvzyh & SN2019gam & 10:19:18.32 & +17:12:42.6 & 0.1235 & ...$^b$ \\
ZTF19aawfbtg & SN2019hge & 22:24:21.20 & +24:47:17.1 & 0.0866 & -19.85 \\
ZTF19acgjpgh & SN2019unb & 09:47:57.02 & +00:49:36.0 & 0.0635 & -20.13 \\
ZTF19abrbsvm & SN2019obk & 22:33:54.08 & -02:09:42.3 & 0.1656 & -19.82 \\
\enddata

\tablecomments{{\it \bf a}: The number is from \citet{DeCia2018}. {\it \bf b}: No data within 100 days around the peak.}
\end{deluxetable}

\end{document}